\documentclass[a4paper,11pt]{article}
\pdfoutput=1
\usepackage{jinstpub} 
\usepackage{subfigure}
\usepackage{graphicx} 
\usepackage{amssymb}
\usepackage{amsmath}
\usepackage{url}
\usepackage{appendix}
\usepackage{booktabs}
\usepackage{float}

\makeatletter
\def\@normalsize{\@setsize\normalsize{10pt}\xpt\@xpt
\abovedisplayskip 10pt plus2pt minus5pt\belowdisplayskip 
\abovedisplayskip \abovedisplayshortskip \z@ 
plus3pt\belowdisplayshortskip 6pt plus3pt 
minus3pt\let\@listi\@listI}

\def\subsize{\@setsize\subsize{12pt}\xipt\@xipt}
\def\section{\@startsection {section}{1}{\z@}{1.0ex plus
1ex minus .2ex}{.2ex plus .2ex}{\large\bf}}
\def\subsection{\@startsection 
   {subsection}{2}{\z@}{.2ex plus 1ex} {.2ex plus .2ex}{\subsize\bf}}
\makeatother
\begin{document}
\date{}

\title{\bf Measurements and Simulations of the Brighter-Fatter Effect in CCD Sensors}

\author{Craig Lage, Andrew Bradshaw, J. Anthony Tyson\\
  Department of Physics\\
  University of California - Davis\\
cslage@ucdavis.edu}

\maketitle

\subsection*{\centering Abstract}

{\em Keywords: 
LSST, modeling, camera, CCD, simulation, diffusion, image processing.  
}
Reduction of images and science analysis from ground-based telescopes such as the LSST requires detailed knowledge of the PSF of the image, which includes components attributable to the instrument as well as components attributable to the atmosphere.  Because the atmospheric component is constantly changing, the PSF is typically extracted from each image by measuring the size and shape of star images across the CCD, then building a fitting function over the focal plane which is used to model the PSF for analysis of extended sources such as galaxies.  Since the stars in each CCD field have a range of brightnesses, accurate knowledge of the PSF for point sources of varying brightness is essential.  It has been found that in thick, fully-depleted CCDs, the electrostatic repulsion of charge stored in the collecting wells gives rise to a larger and slightly more elliptical PSF for brighter stars.  This ``brighter-fatter'' effect has been reported in some detail in the literature.  In this work, we report direct and indirect measurements of this effect in prototype LSST sensors, and describe a detailed physics-based model of the electrostatics and charge transport within the CCD.  

\section{Introduction}

The extended red response of thick fully depleted CCDs enables photometric redshift measurements on high redshift galaxies.  
The thick design with small pixels presents challenges for precision astrometry and weak gravitational lensing   
(the weak lens shear error is the derivative of the astrometric error.)     
Thick fully depleted CCDs exhibit position and intensity dependent charge
transport anomalies which must be understood and corrected in order to reach the
weak gravitational lens shear measurement goals of the next generation probes of dark matter and dark energy, including LSST.  Long confused with QE variations, we must measure, characterize, and correct
for these small scale astrometric variations in the thick fully depleted CCDs. The
effective pixel grid is not exactly rectangular$;$ it is warped. Several effects contribute to
this charge transport redistribution: fringing transverse electric fields around the edge of the chip, 
space charge repulsion of arriving photoelectrons near cores of bright stars, and chemical
potential variations frozen into the silicon.  Many of the small scale features we see in
flat field images actually represent shifted pixel boundaries (and hence shifting pixel
positions and areas).  Historically this has been corrected for with flat-fielding techniques under the assumption that the flat-field variations were due to variations in quantum efficiency.  This essentially introduces a pixel dependent multiplicative factor.  To more correctly account for these effects, we need instead to introduce a remapping of pixel boundaries.  For a given set of voltages, this coordinate mapping is frozen in the chips, so there should be enough
information to constrain the mapping from dithered star array positions.  Correlations in flat fields provide added information, which combined with
the dithered star array data can constrain a device physics model.

The ``brighter-fatter'' (hereafter ``B-F'') effect has been explored by several authors (\cite{antilogus2014}, \cite{doherty2014}, \cite{guyonnet2015}, \cite{gruen2015}, \cite{walter2015}, \cite{rasmussen2015}).  While there have been both empirical measurements of the effect and theoretical treatments, we believe that by building a detailed physics-based model of the CCD, we can help to understand these effects and point the way toward suppressing them through pixel level pipeline image processing techniques. 

Note that the magnitude of the B-F is such that it modifies the sizes of bright images by $\approx 1\%$ or less, and has a progressively smaller impact as images become fainter.  As such, the impact on faint galaxy images can be considered negligible, since the background sky level in ground-based imaging is much larger than the peak galaxy flux.   However, star images, which are used to extract the PSF as a function of position in the focal plane, are significantly impacted and the effect must be corrected for, since their peak intensity is far above the sky level. 

Any such correction to the PSF for the B-F effect will necessarily depend on the brightness of the star as well as its contrast with the background sky and all the camera parameters affecting the B-F effect (voltages, wavelength, temperature). Thus corrections for B-F will have to be made on an exposure by exposure and star by star basis. To do such a correction, a device physics based model of the full anisotropic B-F effect, and its dependence on relevant parameters must be used.    In developing this device physics model of the CCD, it is important to validate the model using lab measurements.

What level of accuracy is required to remove the B-F effect from the stellar images in order to adequtely extract the PSF?  To meet the precision weak lensing science goals of the LSST, it is required to reduce the systematic multiplicative shear bias to the range of $\rm 10^{-3} - 10^{-4}$, and to reduce the additive PSF shear bias to $\rm 10^{-4}$. (\cite{LSST_Science_Book}, \cite{Tyson_Last_Kpc}).  Since the magnitude of the B-F effect is $\approx 1\%$, this means we need to model the effect at the $\rm 1\% - 10\%$ level in order to remove it from the images.

This work is organized as follows. In Section \ref{BF_Meas_Section}, we describe the techniques used for measuring the B-F effect in the lab, both direct measurements of the sizes of artificial stars and indirect measurements of the B-F effect as extracted from pixel-pixel correlations (\cite{antilogus2014}).  In Section \ref{Simulations}, we describe the methods used to simulate the electrostatic effects in the CCD sensor, show some simulated results, and describe how well the model agrees with the measurements.   Finally, we discuss plans for future work and conclude.

\section{Measurements of the Brighter-Fatter Effect}
\label{BF_Meas_Section}
\begin{figure}
\centering
\begin{minipage}{.52\textwidth}
\includegraphics[trim=0.0in 0.0in 0.0in 0.0in,clip,width=\textwidth]{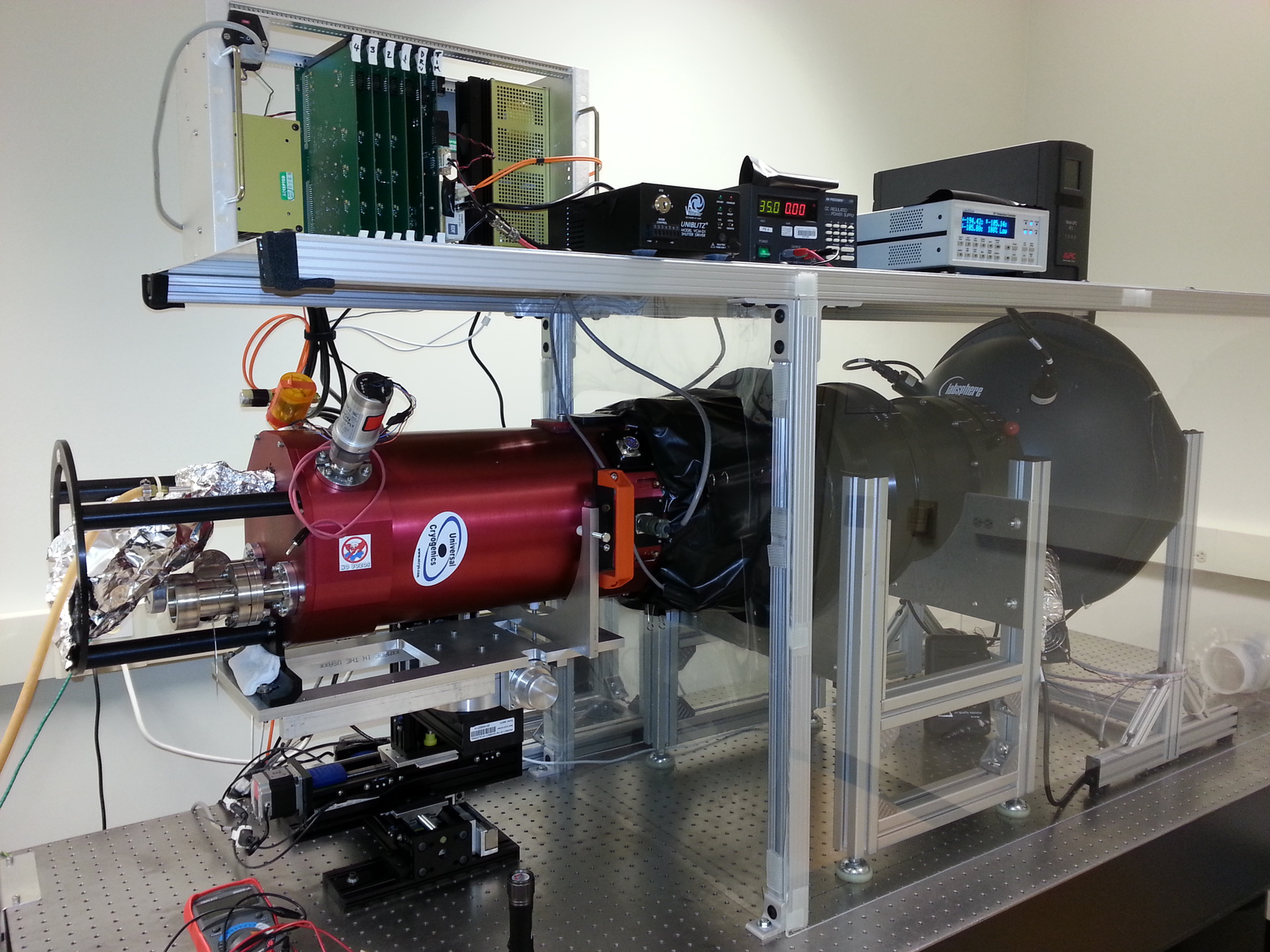}
\end{minipage}
\begin{minipage}{.47\textwidth}
\includegraphics[trim = 1.0in 2.0in 1.0in 2.0in, clip, width=\textwidth]{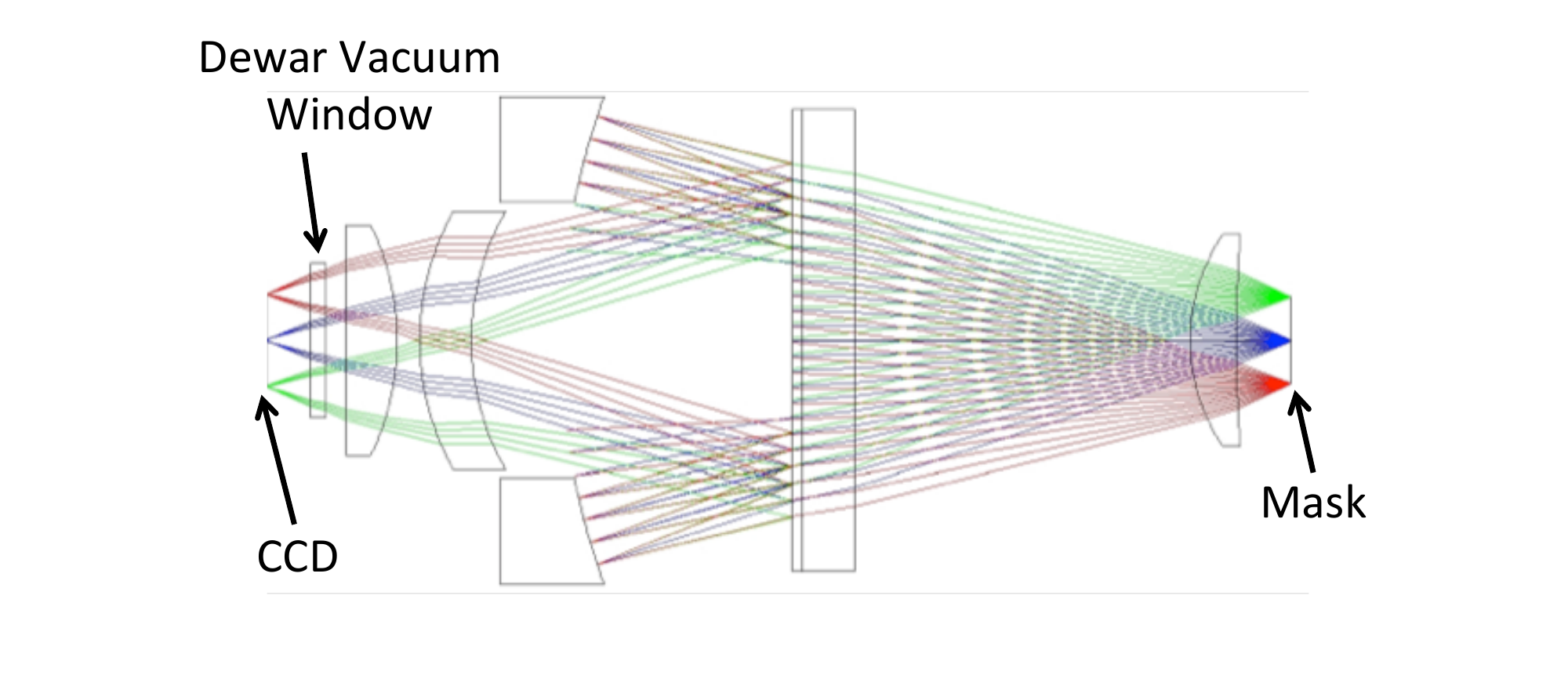}
\includegraphics[trim=0.0in 2.0in 0.0in 2.0in,clip,width=\textwidth]{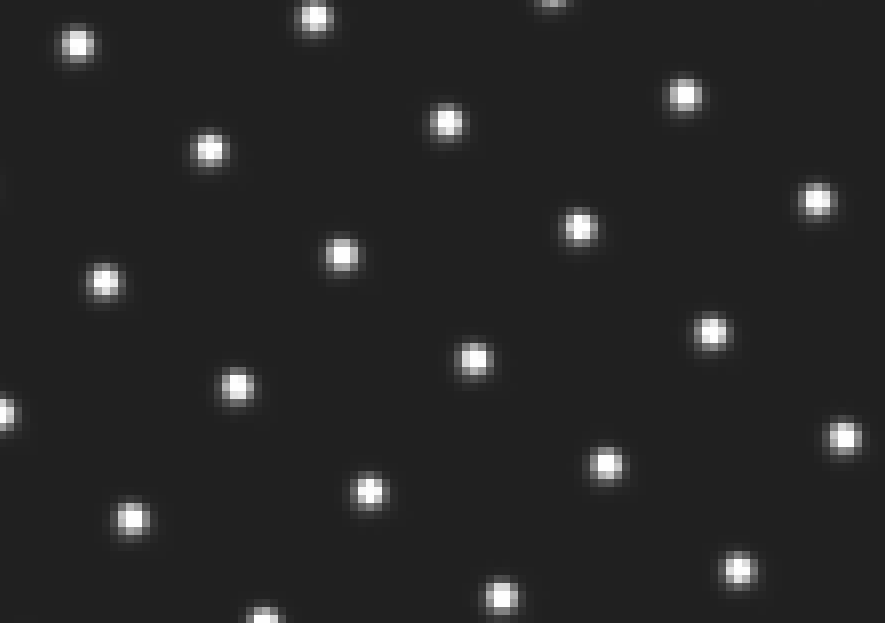}
\end{minipage}
\caption{ The UC Davis LSST Optical Beam Simulator \cite{tyson2014}.  The CCD is contained in the red dewar chamber on the left, the catadioptric optical system is in the center, and the illuminating sphere is on the right.  The control electronics are on top. The upper right image shows the optical system, with the mask on the right, the optical system in the center and the dewar vacuum window and CCD focal plane on the left.  The lower right image is a FITS image of a small portion of the projected array of 30 micron spots.}
  \label{UC_Davis_Simulator}
\end{figure}

Measurements of the B-F Effect are performed using the UC Davis LSST Optical Beam Simulator, which is described in detail in Tyson, et.al. \cite{tyson2014}.  The system, a schematic of the optical path, and a projected image are shown in Figure \ref{UC_Davis_Simulator}.

The LSST Optical Beam Simulator has the capability to
simulate the entire LSST operation, including the centrally obscured  f/1.2 LSST optical beam, a sky with galaxies and
stars with approximately black-body spectra superimposed on a
spatially diffuse night sky emission with its complex emission line features, and provision for spatially dithered imaging.
The system enables a mask containing artificial stars or galaxies to be projected onto the CCD being tested.  Since the CCD currently used is a prototype version of the CCDs intended to be used in the final LSST instrument, and since the optical simulator has been engineered to duplicate the characteristics of the f/1.2 beam in the LSST telescope, it is expected that the results obtained will be representative of those in the final LSST images, without the variable effects of the atmosphere.  In this work, the CCD is an STA3800C prototype CCD produced by the Imaging Technology Laboratory \cite{ITL_website}.  The CCD is comprised of 16 segments, each containing 4000x509 pixels which are 10 microns square, for a total imaging area of 4000x4072 pixels or 40mm x 40.72 mm.  In the final LSST instrument, the image scale is such that each pixel will have an angular size of 0.2 arcseconds.  The CCD has a thickness of 100 microns of fully-depleted silicon in order to enhance the quantum efficiency in the near infrared (up to $\rm \lambda = 1.05 \mu m$) region.

Two different techniques are used to characterize the B-F effect on the optical simulator.  The first is direct imaging of artificial stars of differing intensities at many dithered positions.  The second is measurement of flat field images and extraction of pixel-pixel correlations, as described in \cite{antilogus2014}.  The details of these measurements are described in the next sections.
\vspace{2mm} 
\subsection{Direct Measurements of Spot Sizes}
\label{Spot_Sizes}
Using the LSST Optical Beam Simulator, we are able to directly measure the B-F effect on arrays of artificial stars (i.e.''spots'').  To generate the spot images, a mask is placed at the focal plane of the optical simulator (on the right in Figure \ref{UC_Davis_Simulator}).  The mask consists of a 2mm thick glass plate with an opaque metallic layer,  which has been patterned with an array of $\rm \approx 40000$ holes, each of which is about 30 microns in diameter \cite{bradshaw2015}.  The 30 micron size has been chosen to give images approximately 0.6 arcseconds in diameter, simulating the expected conditions at the LSST on nights of good seeing.  The mask is illuminated using a scattering sphere which illuminates the mask with a uniform intensity.  The 1:1 f/1.2 re-imaging optics generates an image focused onto the CCD under test.  For all of these tests, an LSST r-band filter is used (center wavelength of 621 nm), and the CCD is maintained at $\rm -100^{\circ}C$ through the use of an LN2 reservoir and a feedback-controlled resistive heater.  Additional investigations at longer wavelengths will be reported elsewhere.

The image of a single spot contains too much shot noise to allow determination of the spot size to the level of accuracy needed to characterize the B-F effect, which typically modifies the spot sizes by 1\% or less.  In order to reach the required level of accuracy, we image a large number (typically 500-1000) of spots simultaneously for each region of the CCD, and then use forward modeling techniques to extract the best estimate of the spot size and shape.  A brief description of the modeling sequence is as follows: 
\begin{enumerate}
  \itemsep 0.05mm
  \item Use Sextractor \cite{Bertin_sex} to identify a list of spots.
    \begin{enumerate}
      \itemsep 0.05mm
      \item Use Sextractor central pixel location.  Note that Sextractor is only used to identify the pixel containing the largest signal, and the spot size and spot central location within the pixel as determined by Sextractor are not used.
      \item Use a constant window (``postage stamp'') for all spots.  Here we are using 9x9 pixels.
    \end{enumerate}
  \item Assume all spots in a local region have the same shape, but allow variable peak intensity and offset within central pixel. 
    \begin{enumerate}
      \itemsep 0.05mm
      \item Calculate first moment of postage stamp to determine offset within central pixel.
    \end{enumerate}
  \item Assume a 2D elliptical Gaussian, calculate expected signal in each pixel.
  \item Find $\rm (\sigma_x, \sigma_y)$ which minimizes:
      $\rm \sum_{Nspots}\sum_{x,y}(Measured_{n,x,y} - Calculated_{n,x,y})^2$
\end{enumerate}

The forward modeling routine was tested on both measured and mock data to ensure it gives consistent results.  These results were quite satisfactory, and gave confidence in using this technique to characterize the B-F effect. 

To characterize the B-F effect, we next apply the above techniques to images generated with a range of photon fluxes.  We do this by keeping the light intensity fixed and increasing the exposure time to increase the spot flux. Figure \ref{BF_Meas} (a) shows a typical result.  Shot noise dominates at low flux levels, and saturation effects set in at high flux levels, but there is an intermediate region where the spot size increases linearly with flux.  The slope of the line in this region is used to quantify the B-F effect.  Note that the slope of the line in the Y-direction (parallel to the CCD channel stops) is always greater than the slope in the X-direction (parallel to the CCD gates).  This is because the electric field  near the ``frontside'' readout gate structure due to the channel stops which confines the collected charge is larger than the confining electric field due to the CCD parallel gates.  This causes the spots to become slightly elliptical in the Y-direction as the flux increases, and is an important effect to model because it will have a systematic effect on the measurement of galaxy shapes and consequently on measurements of weak lensing.
\vspace{2mm} 
\subsection{Pixel-Pixel Correlation Measurements}
\begin{figure}
\begin{center}
  \subfigure[\large{Direct measurements of spot size increase.}]{\includegraphics[trim=0.0in 0.0in 0.0in 0.5in,clip,width=0.51\textwidth]{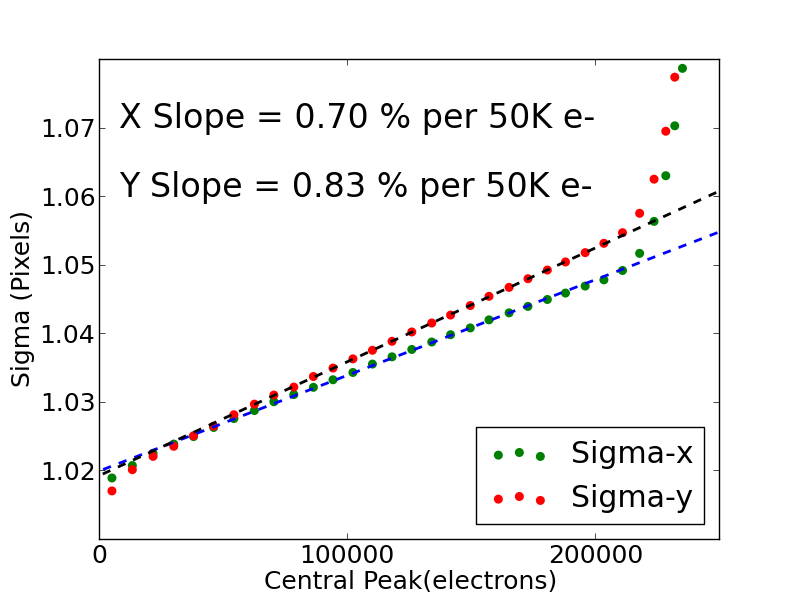}}
   \subfigure[\large{Pixel-pixel correlations.}]{\includegraphics[trim=0.1in 0.0in 0.5in 0.5in,clip,width=0.48\textwidth]{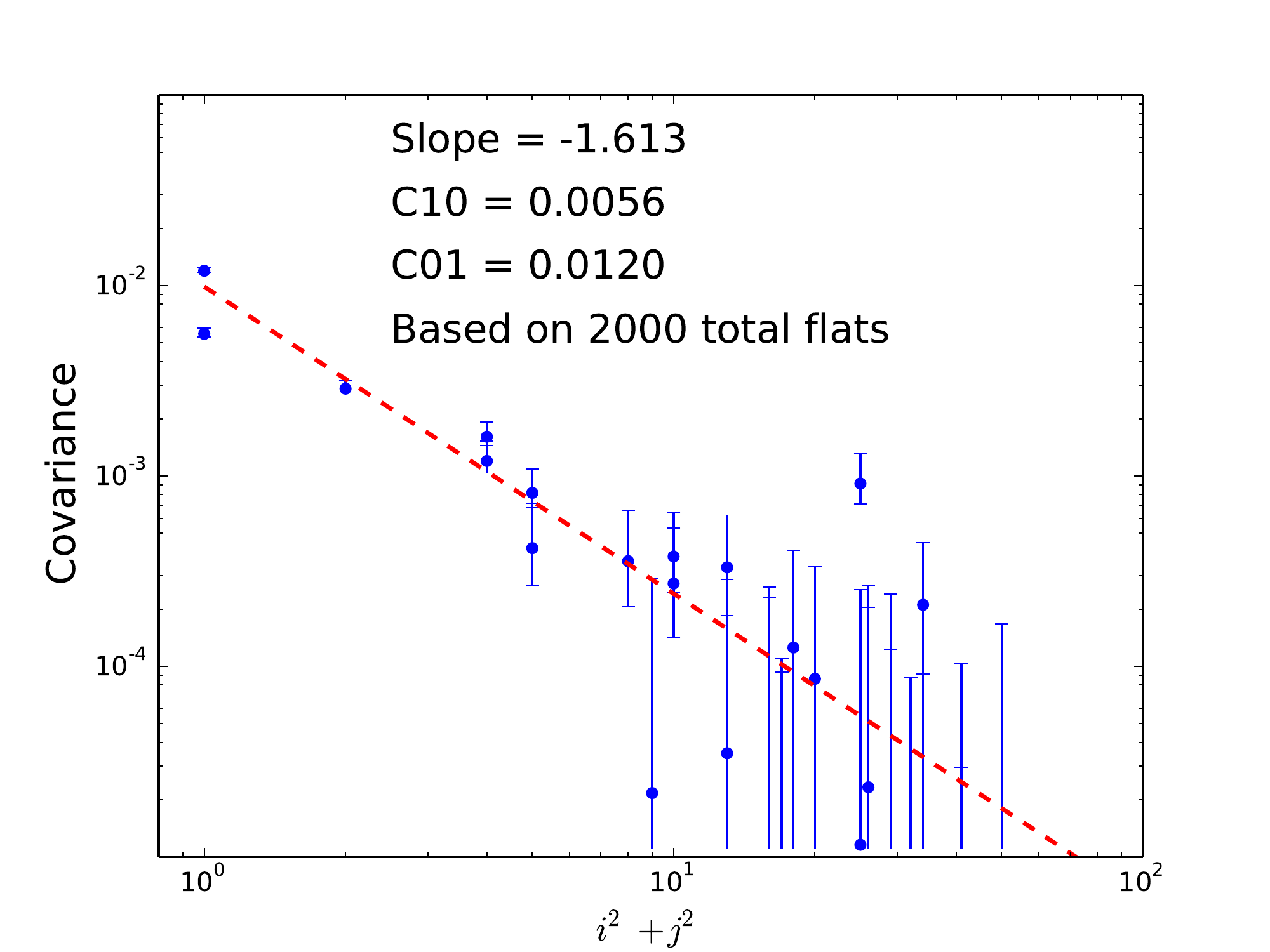}}
        \caption{Figure (a) shows measured spot sizes as a function of the average stored charge level in the central (peak) spot. The Y-direction is parallel to the CCD channel stops, and the X-direction is parallel to the CCD parallel gates.  Note that the spots become both larger and more elliptical as the flux increases, and become rapidly larger as saturation sets in at high flux.  Figure (b) shows measurements of pixel-pixel covariances taken by averaging 1000 pairs of flats at each flux level, and extrapolating to a value of 80,000 electrons in the central pixel.  The dashed red line is a best fit slope, showing the covariance decreasing as the distance increases.
\label{BF_Meas}}
\end{center}
\end{figure}

A second method of characterizing the B-F effect, as originally described by Antilogus, et.al. \cite{antilogus2014}, involves measurements of pixel-pixel correlations averaged over large numbers of flat images.  If, through random variation, a pixel accumulates more charge than the average,  electrons are repelled from this pixel into surrounding pixels.  This causes positive correlations with surrounding pixels, since an increase in charge of a given pixel causes a corresponding increase in surrounding pixels due to space charge repulsion.  By measuring and averaging large numbers of flats, these correlations can be seen and quantified.  However, there are several subtleties to the analysis of this data.  First, because the light intensity of the flats is usually not perfectly uniform, it is best to subtract pairs of flats before calculating the pixel-pixel correlations.  After subtracting pairs of flats, we are left with a pixelized difference image from which the correlations are calculated.  The correlations are divided by the variance in order to yield a covariance.
There are other effects, such as time correlated electronics noise, that can cause pixel-pixel correlations, and these must be separated from the correlations due to the B-F effect.  To do this, we calculate the covariances at a range of flux levels and extrapolate back to zero flux.  The intercept at zero flux is then subtracted off to identify the part of the correlation which is proportional to flux level, and the covariance at a given flux level (typically 80,000 electrons) is calculated.  Figure \ref{BF_Meas} (b) shows a typical result of this procedure.  Both the direct spot size measurements and the pixel-pixel correlation measurements have been repeated multiple times at a variety of measurement conditions, and on 12 different segments of the CCD sensor.  Section \ref{Meas_Sim} will discuss how these measured results compare to simulations of the charge transport in the CCD.

\section{Simulations of the Brighter-Fatter Effect}
\label{Simulations}
As part of this work, we have built a detailed physics-based model of the electric field and charge transport within the CCD, and compared the simulations which result to the measurements described in the last section.  This simulation code is freely available at \cite{CCD-Code}, and will be described in detail in a separate publication.  The next two sections give an overview of the simulation methods.
\vspace{2mm} 
\subsection{Solving for the Electric Field} 
\begin{figure}
  \centering
  \subfigure{\includegraphics[trim=0.0in 0.0in 0.0in 0.1in,clip,width=0.43\textwidth]{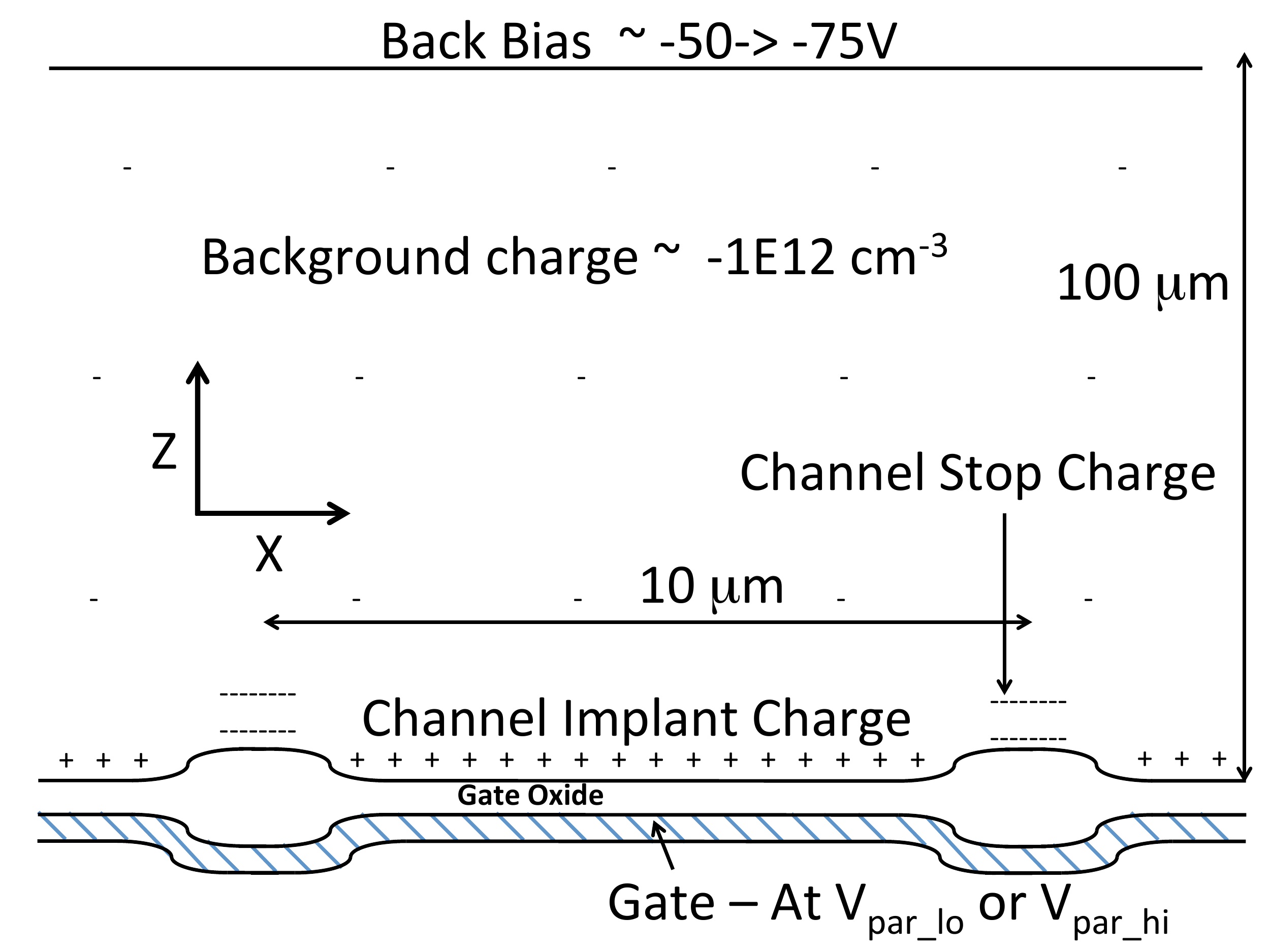}}
  \subfigure{\includegraphics[trim=0.0in 0.0in 0.0in 0.1in,clip,width=0.43\textwidth]{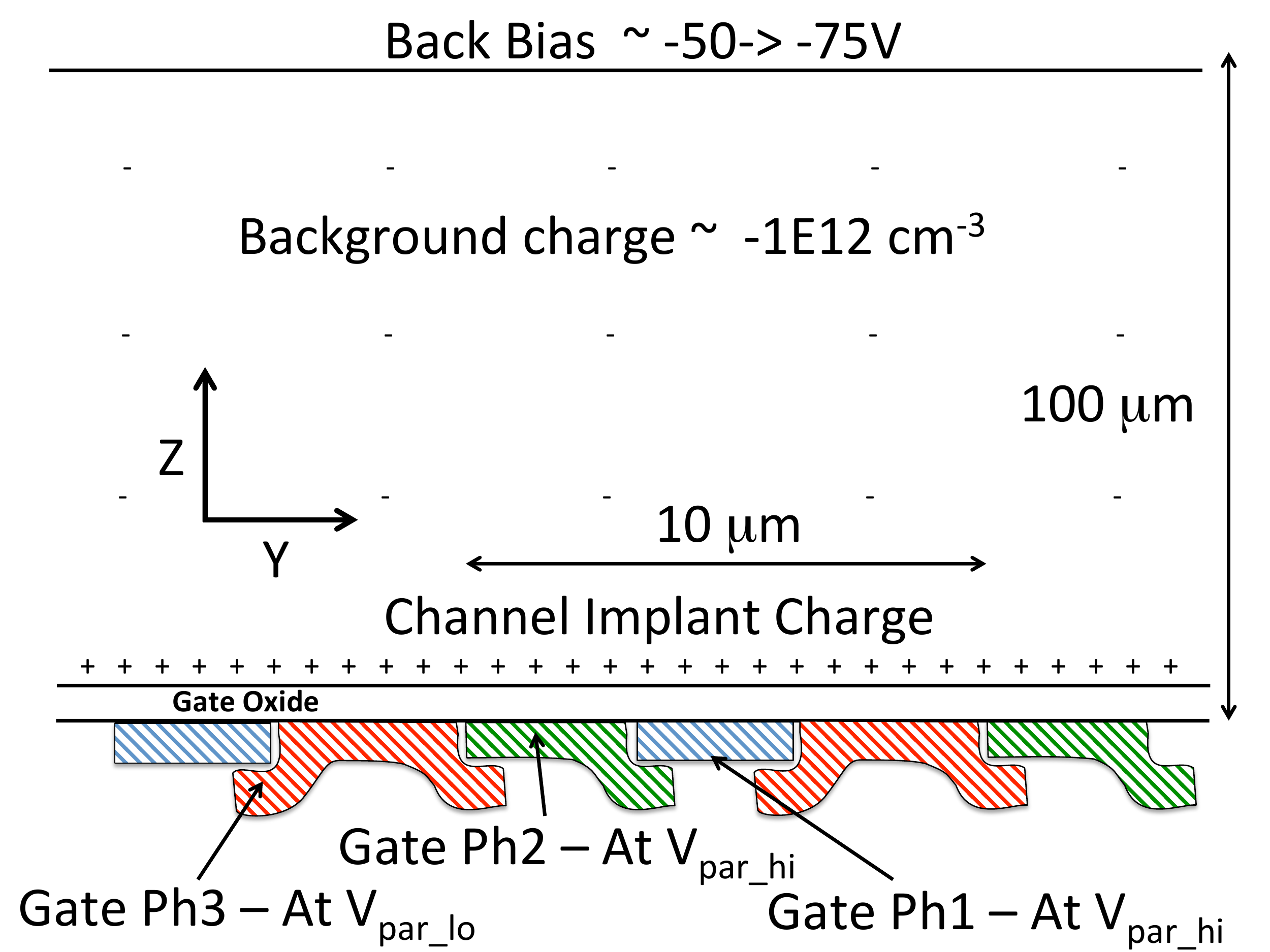}}
  \caption{Basic cross section of the CCD structure.  Here light is incident at the top of the diagram and photoelectrons are collected at the bottom.  The left diagram is parallel to the CCD parallel gates, and the right diagram is perpendicular to these gates.  Note the difference in vertical and horizontal scales.}
  \label{CCD_Cross_1}
\end{figure}

\begin{figure}
  \centering
  \includegraphics[trim=0.0in 3.0in 0.0in 3.0in,clip,width=0.80\textwidth]{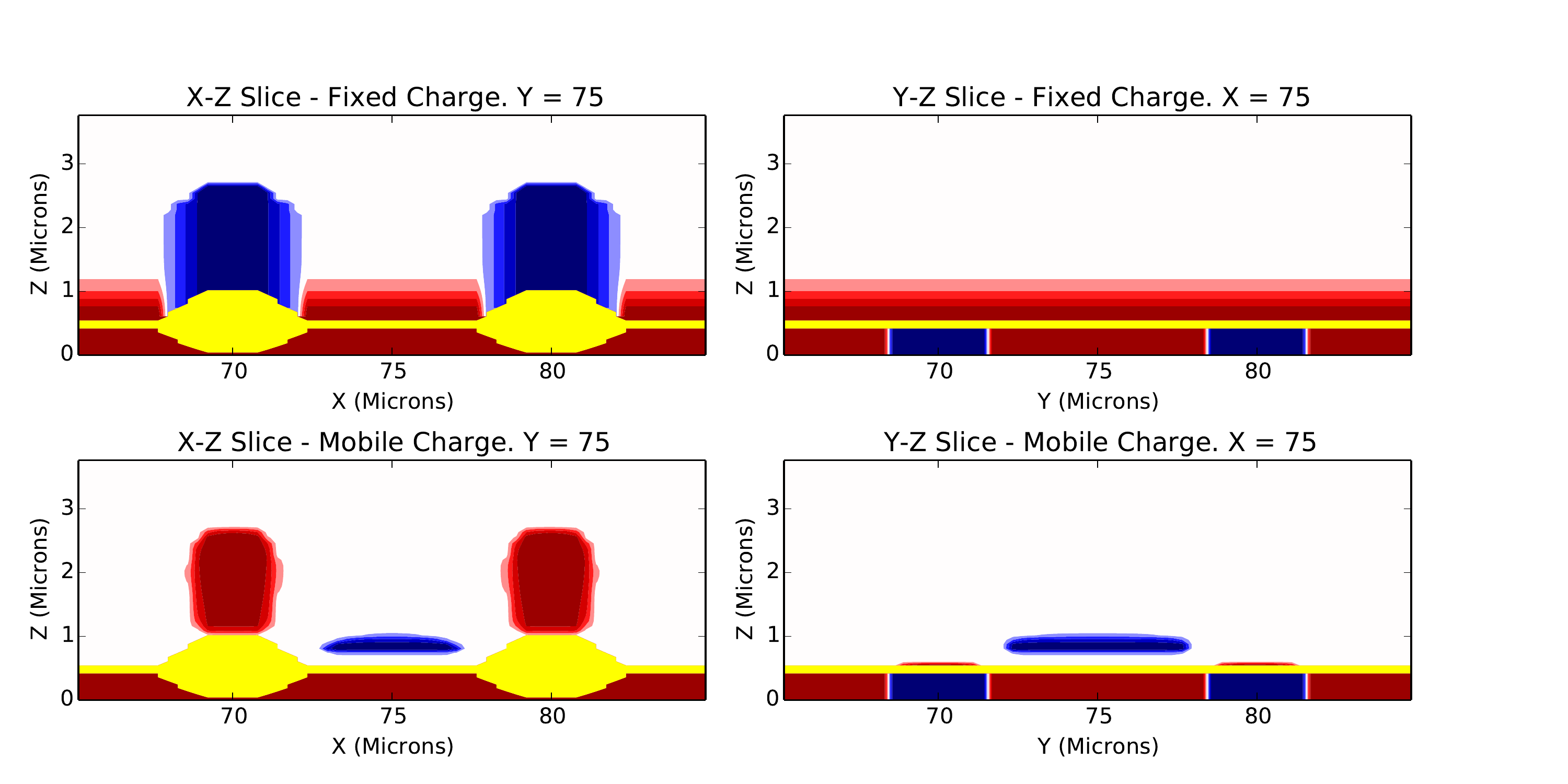}
  \caption{  These cross-sections are from the simulation and show the fixed and mobile charge densities, respectively, with positive charges in red and negative charges in blue.  Note the large density of free holes in the channel stop region (lower left), as well as the free holes present in an inversion layer under the negatively biased barrier gates in the channel region (lower right).  This simulation has 80,000 electrons stored in the central pixel with the surrounding pixels empty.  Note the difference in vertical and horizontal scales.}
  \label{CCD_Cross_2}
\end{figure}

\begin{figure}
  \centering
  \includegraphics[trim=0.0in 0.0in 0.0in 0.4in,clip,width=0.99\textwidth]{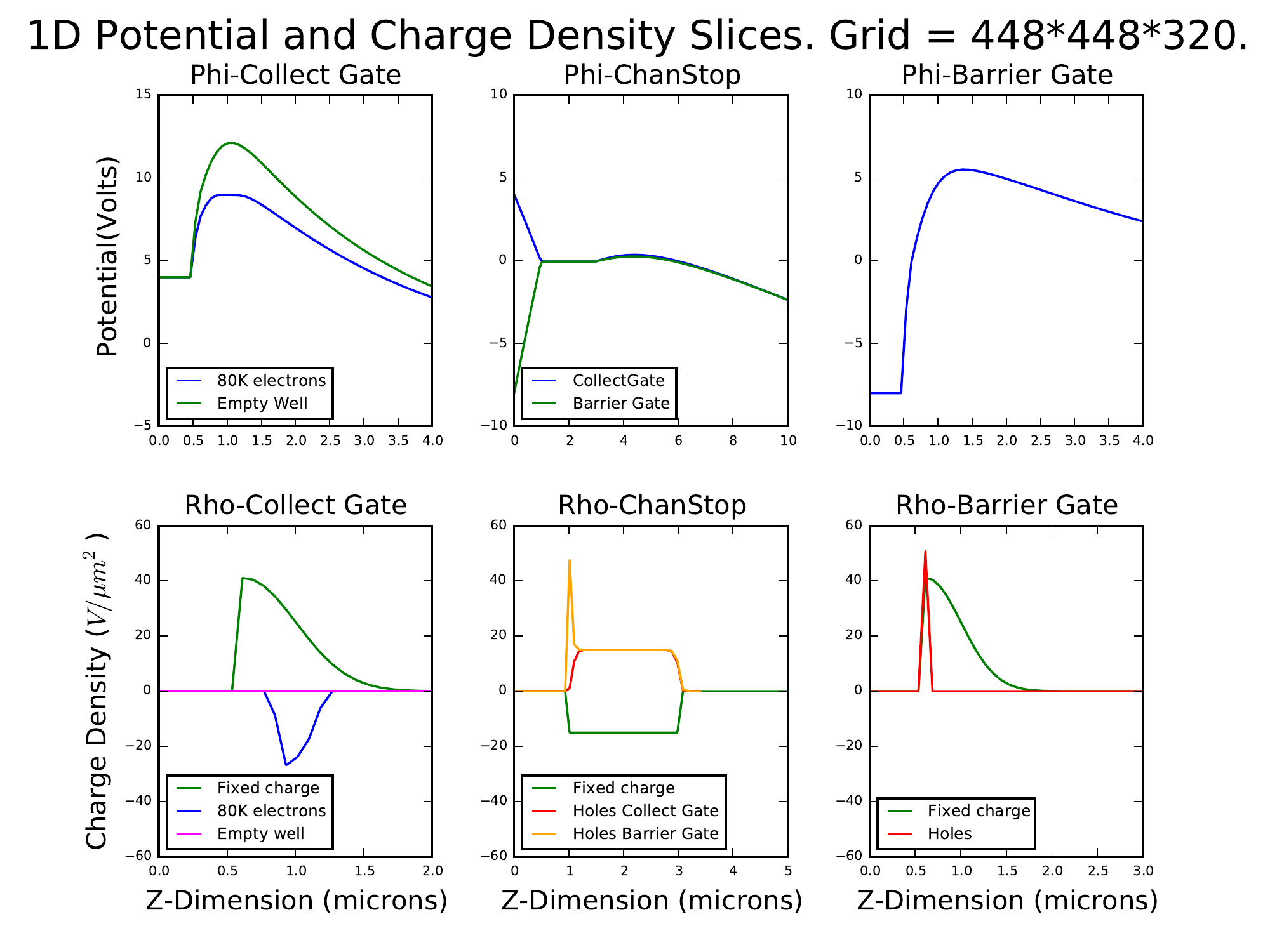}
  \caption{One dimensional profiles in the Z-direction of the Poisson's equation solution for conditions which best reproduce the measured results.  Top Row: Potentials in the center of a collecting gate ($\rm V_{parallel-hi}$), the center of the channel stop region, and the center of a barrier gate($\rm V_{parallel-lo}$), respectively.  Bottom Row: Charge densities in these same regions (charge densities are given as $\rm \rho/\epsilon_{Si}$ in units of $\rm V/\mu m^2$).  The collecting gates show one empty pixel and one pixel containing 80K electrons.  In the top center plot, note the potential barrier at $\rm Z \approx 5\mu m$ which prevents the free holes in the channel stop region from flowing to the negatively biased backside of the CCD.} 
  \label{Potential_1D}
\end{figure}

Figures \ref{CCD_Cross_1} and \ref{CCD_Cross_2} show cross sections of the basic CCD structure.  An excellent overview of the physics of fully depleted CCDs is given by Holland, et.al. \cite{holland2014}.  The incident photons create hole-electron pairs in the silicon, and a large electric field is set up so that positively charged holes are swept to the back side and negatively charged electrons are swept to the front side where the polysilicon gates reside.  The electrons are collected under the polysilicon gates which define the pixels of the device, where they are read out during clocking of the collecting gates.  The CCDs in this work are fully depleted p type silicon which is $\rm 100 \mu m$ thick, and pixels which are $\rm 10 \mu m$ square, which is a vertical to horizontal ratio of 10:1.  Ideally the electrons propagate purely vertically so that a photon incident in a given pixel is collected in the same pixel.  However, transverse electric fields can cause electrons to propagate into pixels offset from the incident photon, resulting in distortions of astronomical images.  In order to simulate this process, it is necessary to first simulate the electric field in the silicon.  This is done by solving Poisson's equation within a discretized volume given the potentials on the boundary of the volume and the charges within the silicon volume, using multi-grid methods.  The potentials on the boundaries are known given the voltages applied to the device.  The fixed charge densities are defined by the dopant profiles introduced when the device is manufactured.  These are not known precisely, but estimates are made and these profiles are then varied until there is a good fit between the measured and simulated results.  Calculating the mobile charge densities is the most difficult.  These are calculated by assuming that regions containing mobile charges are in quasi-equilibrium, and setting a quasi-Fermi level for each region containing mobile charge.  We find that there is a significant density of free holes, both in the channel stop region, and in an inversion layer under the negatively-biased barrier gates.   This cross-matrix of free holes is referenced back to ground potential in our device, so a single quasi-Fermi level for holes is set throughout the device to maintain this region near zero volts.  For the electrons, the quasi-Fermi level in each pixel is iteratively adjusted to give the appropriate number of electrons in that pixel.  Note the location of the free holes and the stored electrons in Figures \ref{CCD_Cross_2} and \ref{Potential_1D}.  

Figure \ref{Potential_1D} shows plots of the charge densities and potential solution for a set of device geometries which best reproduces the measured results.  The upper left of Figure \ref{Potential_1D} clearly shows the buried channel region generated by the channel implant.  This creates a potential well displaced from the $\rm Si-SiO_2$ interface where the electrons are stored.  Since recombination velocities at this interface are much higher than in the bulk silicon, storing the charge away from the interface greatly increases the charge lifetime and charge transfer efficiency.  Also note the presence of free holes in the channel stop region, and the surface inversion layer of free holes present under the negatively biased barrier gates.  The top center potential plot also shows the potential barrier which prevents the free holes in the channel stop region from flowing to the negatively biased backside of the CCD.

\vspace{2mm} 
\subsection{Simulating the Charge Transport}
\begin{figure}
\begin{center}
\subfigure[]{\includegraphics[trim=0.0in 0.0in 0.0in 0.0in,clip,width=0.30\textwidth]{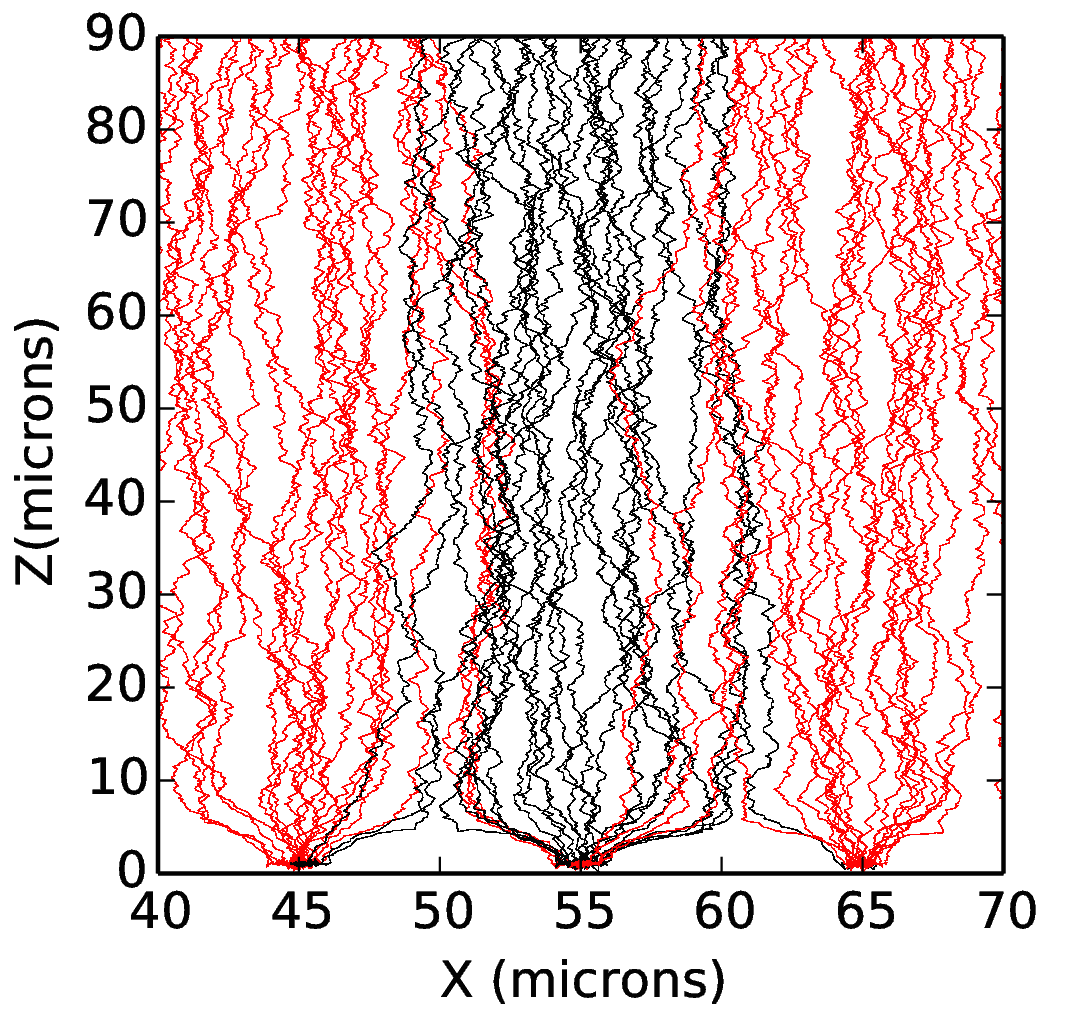}}
\subfigure[]{\includegraphics[trim=0.5in 1.2in 4.0in 1.5in,clip,width=0.30\textwidth]{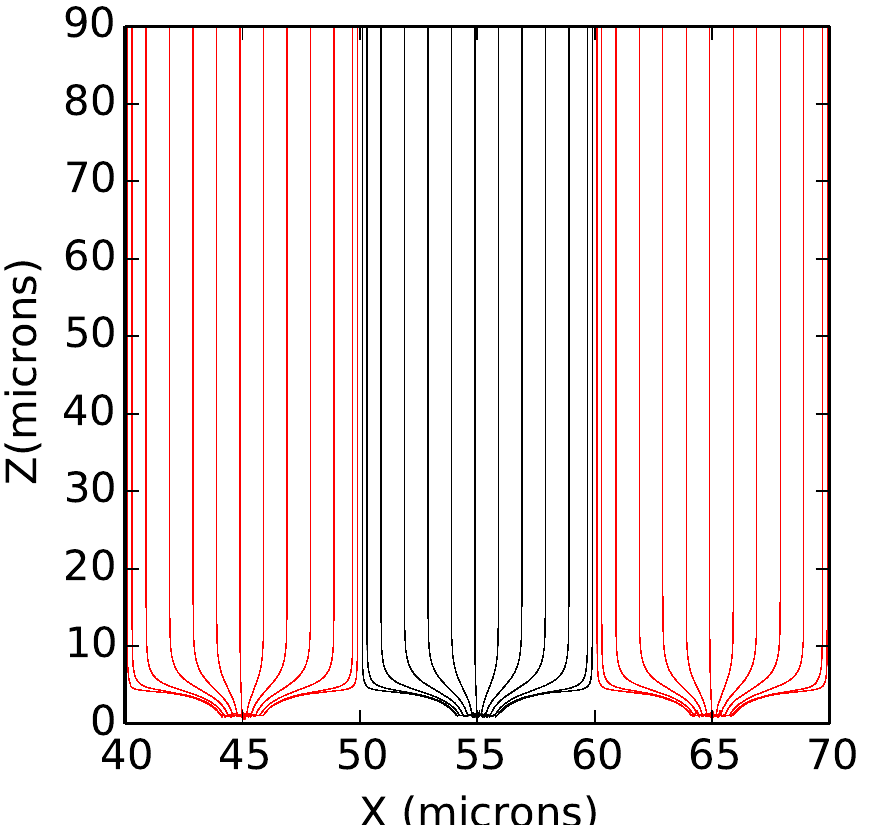}}
\subfigure[]{\includegraphics[trim=0.50in 1.2in 4.0in 1.5in,clip,width=0.30\textwidth]{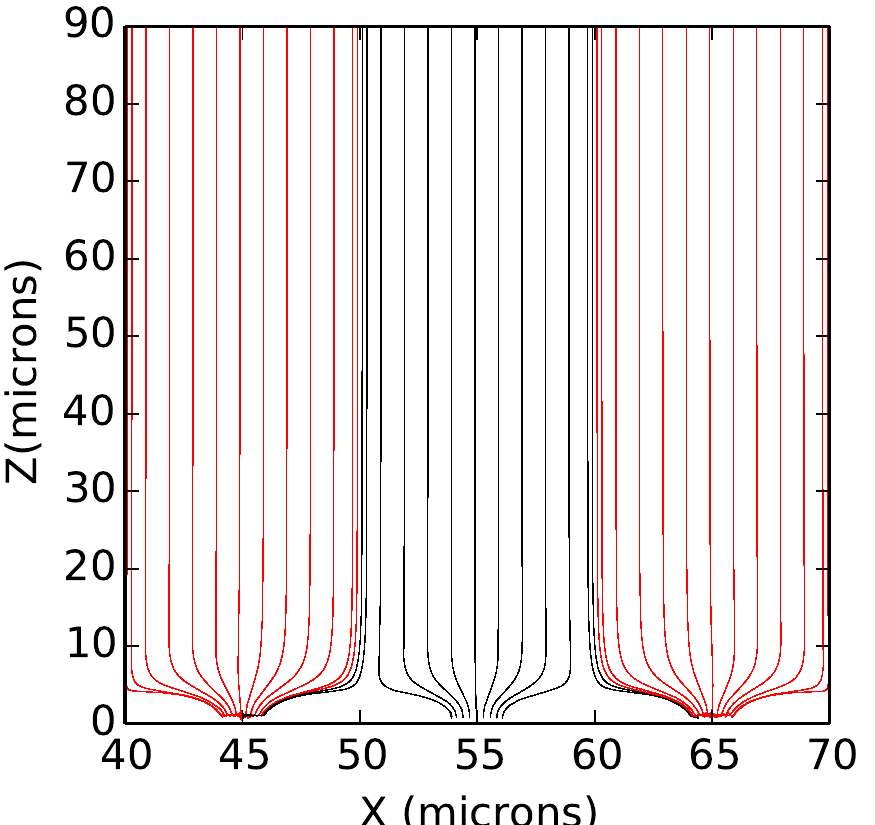}}

\end{center}
\caption{Impact of electron diffusion. (a) Theoretical diffusion. (b) No diffusion. (c) No diffusion with the central pixel containing 160,000 electrons, showing electrons being repelled into surrounding pixels.  Electron paths are color coded depending on within which undistorted pixel they were created.}
\label{Electron_Diff}
\end{figure}

Once we have a solution for the potential and electric field within the silicon, the next step is to calculate the electron paths, so we can simulate where the photoelectrons created by the incident photons are ultimately stored.  This is done by simulating repeated scattering steps of the electron, with the time between scattering events being drawn from an exponential distribution of the known scattering time, and with the electron motion between scattering events being a superposition of the drift motion in the direction of the electric field and a thermal motion in a randomly chosen direction.  Since the diffusion coefficient and mobility of electrons in silicon are well known, there are no free parameters in this simulation.  Figure \ref{Electron_Diff} shows the impact of diffusion on the electron paths and on the sharing of electrons between pixels.  

The presence of stored charge in a pixel repels incoming electrons, pushing them into surrounding pixels; this is the cause of the B-F effect.  Figure \ref{Electron_Diff} (c) shows the impact of one pixel containing 180K electrons on the surrounding electron paths.

Using the simulations of the electron paths, one can also study the distortion of apparent pixel boundaries due to the B-F effect.  To do this, we use a binary search to locate the pixel boundaries, with results as shown in Figure \ref{Pixel_Distortion}.  To make this plot, we first solve Poisson's equation with the central pixel containing charge (here 80,000 electrons).  Then, with diffusion turned off to eliminate the stochastic effects, we propagate electrons down to the collecting wells to determine in which pixel they end up.  We perform a binary search along a radial path starting at the pixel center until we have located the pixel boundary to a desired level of precision (here 0.001 micron).  This allows us to determine the changes in the effective pixel boundaries and in the pixel areas.  For these plots, each pixel polygon is delineated by 132 distinct vertices.  The change in pixel area allows us to predict the measured pixel-pixel correlations, as will be discussed in the next section.

\subsection{Simulating the B-F Effect and Comparisons of Measurements with Simulations}
\label{Meas_Sim}
We next discuss applying the simulation machinery discussed in the last sections to the B-F effect.  The strategy for simulating the direct spot size B-F slopes and for simulating the pixel-pixel correlations is somewhat different.  To simulate the direct spot size B-F slope measurements, as seen in Figure \ref{BF_Meas} (a), we directly simulate an array of spots very similar to the array of spots that were measured to produce Figure \ref{BF_Meas} (a).  The array of simulated spots is produced as follows:
\begin{enumerate}
  \itemsep 0.05 mm
  \item Solve Poisson's equation for a postage stamp (using 13x13 pixels) with all pixels empty.
  \item Choose a random location within the central pixel as the center of a 2D Gaussian spot.  To match the measured 30 micron spots discussed in Section \ref{BF_Meas_Section}, the Gaussian spot has $\rm \sigma \approx 10$ microns.
  \item Determine starting locations for N electrons in the 2D Gaussian spot.
  \item Propagate these electrons down to their collecting gates.  It typically takes $\approx 1000$ diffusion steps for the electrons to travel from the creation point down to the collecting gate.  After the electron has reached a storage location, we allow it to equilibrate for some period of time (typically 1000 diffusion steps), tracking these locations to simulate the equilibrium charge density.
  \item Re-solve Poisson's equation with these wells now containing the appropriate charge.
  \item Repeat with N more electrons.  Perfect accuracy would demand N = 1, but the electric field changes only slightly with $\rm <1000$ electrons / pixel.  We have been using N=10,000 which places about 1000 electrons in the central pixel, so about 100 iterations are needed to fill the central pixel.
  \item Repeat for more than one spot (typically $\rm \approx 64$), each with a different random center displacement.  Multiple spots can all be simulated in parallel.
  \item Use the resulting array of spots as input to the same forward modeling photometry routine used to analyze the measured data (see Section \ref{BF_Meas_Section}).
\end{enumerate}

\begin{figure}
  \begin{minipage}{0.34\textwidth}
  \includegraphics[trim=1.4in 0.4in 1.0in 0.3in,clip,width=1.10\textwidth]{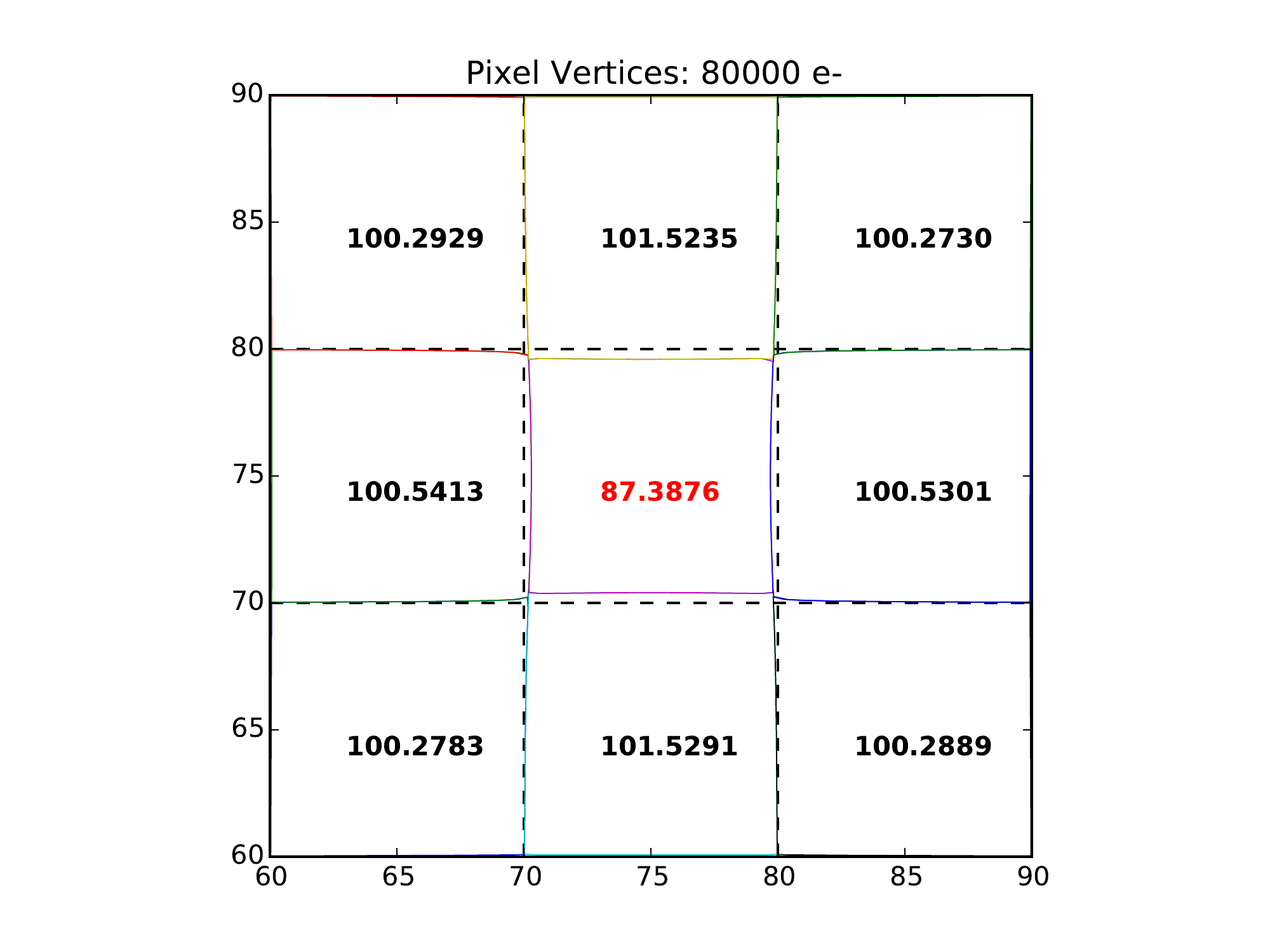}
  \caption{Distortion of pixel boundaries and areas due to 80,000 electrons in the central pixel.  Undistorted pixel areas are 100 square microns. Each pixel polygon in this diagram contains 132 vertices.  The vertices are found through binary search.  Diffusion is turned off to make this map.  These distorted pixel areas are used to calculate the expected pixel-pixel correlations, as explained in the text.}
  \label{Pixel_Distortion}
  \end{minipage}
  \hspace{5mm}
  \begin{minipage}{0.64\textwidth}
    \includegraphics[trim=0.0in 0.8in 0.0in 0.8in,clip,width=0.99\textwidth]{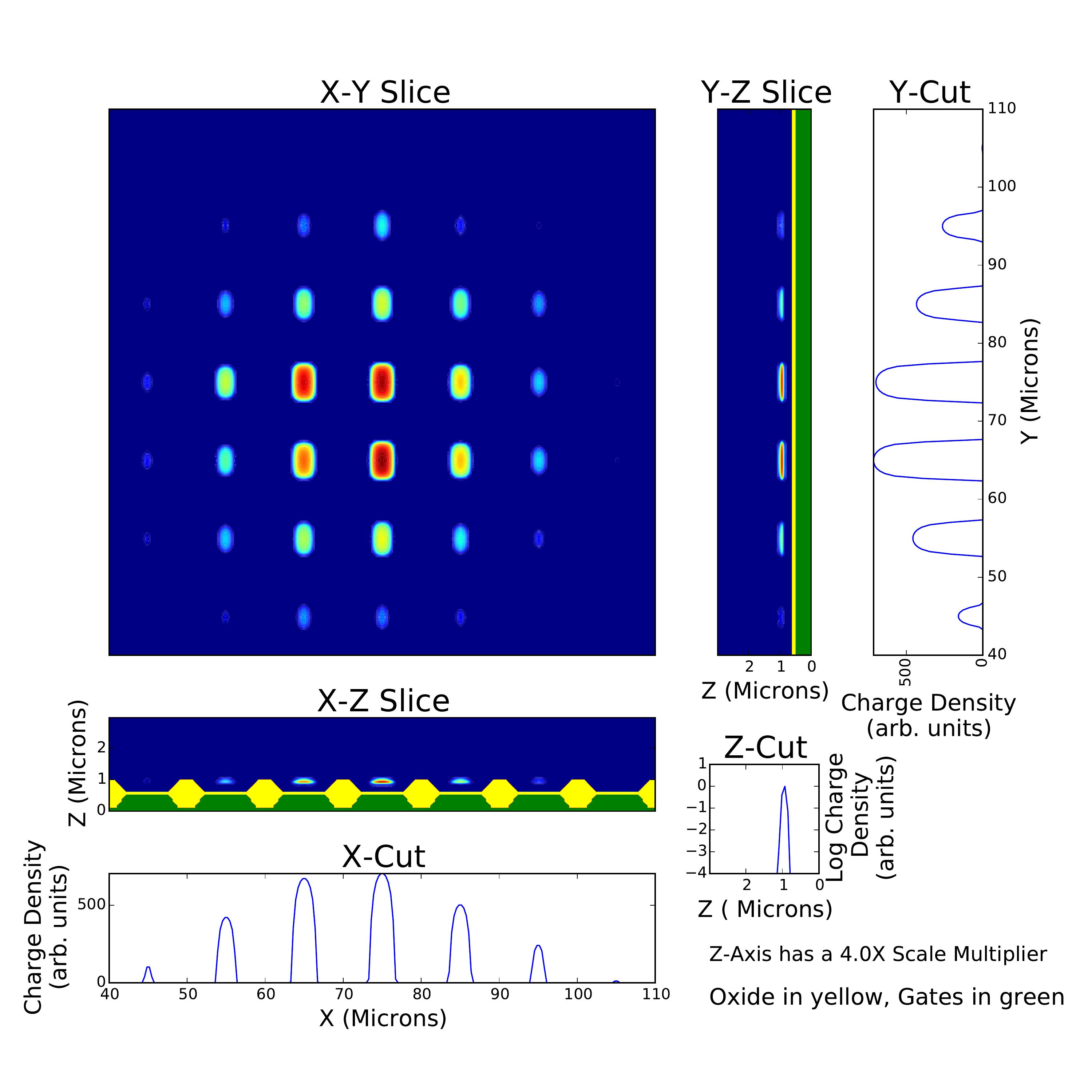}
    \caption{This plot shows the accumulation of charge in a typical spot as the B-F simulation progresses.  This spot has accumulated 200,000 electrons in a Gaussian spot with a 10 micron sigma, which places approximately 20,000 electrons in the central pixel. This particular spot has the spot center displaced down by 4.51 microns and to the left by 2.75 microns.}
\label{Charge_Buildup}
  \end{minipage}
\end{figure}

We can track the buildup of charge in the collecting wells as the simulation progresses.  Figure \ref{Charge_Buildup} shows this progression.  

The measurements of pixel-pixel correlations, as shown in Figure \ref{BF_Meas} (b), can also be calculated from the simulation.  Directly simulating a series of flats like the ones used to generate Figure \ref{BF_Meas} (b) is computationally out of the question, as the flats used to produce this figure contain $\rm \approx 10^{15}$ photons.   However, as seen in Appendix \ref{Correlations_Appendix}, it can be shown that for small area deviations the pixel-pixel correlations are determined by the distortions in the pixel areas.     So all that is required is a single simulation with N electrons (we use here N = 80,000) in the central pixel, followed by determining the distorted pixel areas of all of the surrounding pixels, as shown in Figure \ref{Pixel_Distortion}.    One then compares a measurement of correlations where each pixel has an average of N electrons, to the simulated area change due to a central pixel with N electrons while the surrounding pixels are empty (as in Figure \ref{Pixel_Distortion}).    Figure \ref{Meas_vs_Sim} shows the comparison of measurements and simulations for several different voltage configurations.   There is good agreement between the measurements and the simulation for both the direct B-F slopes (left column) and the pixel-pixel correlation measurements (right column).  The fit to the B-F slopes is within the error of the measurements, and is also within the 1-10\% range which is needed to adequately remove the impact of the PSF systematic errors from multiplicative and additive shear bias in weak lensing measurements.  We believe the concordance can be improved, and the number of marginalized parameters can be reduced,  by having direct physical measurements of the CCD (especially oxide thicknesses and dopant profiles).

\section{Conclusions and Next Steps}
\label{Conclusions}
In this work we have described a physics-based charge transport simulation of the CCD which accurately models the B-F effect, as determined by both direct measurements of spot sizes and by measurements of pixel-pixel correlations in CCD flats, both as a function of flux and voltages.  The model is capable of simulating the B-F effect in PSF calibration stars to the accuracy required for determining the PSF in LSST images and suppressing the systematic impact of this effect on measurements of galaxy shapes to the level required by the weak lensing science.  We believe the goodness of fit between the model and the simulation can be improved, and the model can be made more predictive, by obtaining physical measurements of the CCD being tested (essentially oxide thicknesses and dopant profiles).  These measurements, as well as quantifying the variation from CCD to CCD, are planned for the future.  Additional measurements are also being taken as a function of wavelength and temperature to further extend the range of validity of the model.  Another area of focus is to incorporate the results of these simulations into faster image simulators, such as GalSim (\cite{rowe2014}, \cite{galsim-code}).  This is being done by incorporating the distorted pixel shapes, as seen in Figure \ref{Pixel_Distortion}, into the image simulator.

\section{Acknowledgements}
We thank Kirk Gilmore for significant help with the CCD readout electronics and dewar hardware.  Many thanks to Perry Gee for lab and IT support.  We received helpful suggestions from Pierre Astier, Andrei Nomerotski, Andy Rasmussen, John Peterson, Rachel Mandelbaum, Michael Schneider, and Debbie Bard.  Financial support from DOE grant DE-SC0009999 and Heising-Simons Foundation grant 2015-106 are gratefully acknowledged.  Computing resources for this work were provided in part by LLNL Laboratory Directed Research and Development grant 16-ERD-013.

\begin {figure}[H]
  \centering
  \subfigure{\includegraphics[trim=0.2in 0.1in 0.5in 0.6in,clip,width=0.495\textwidth]{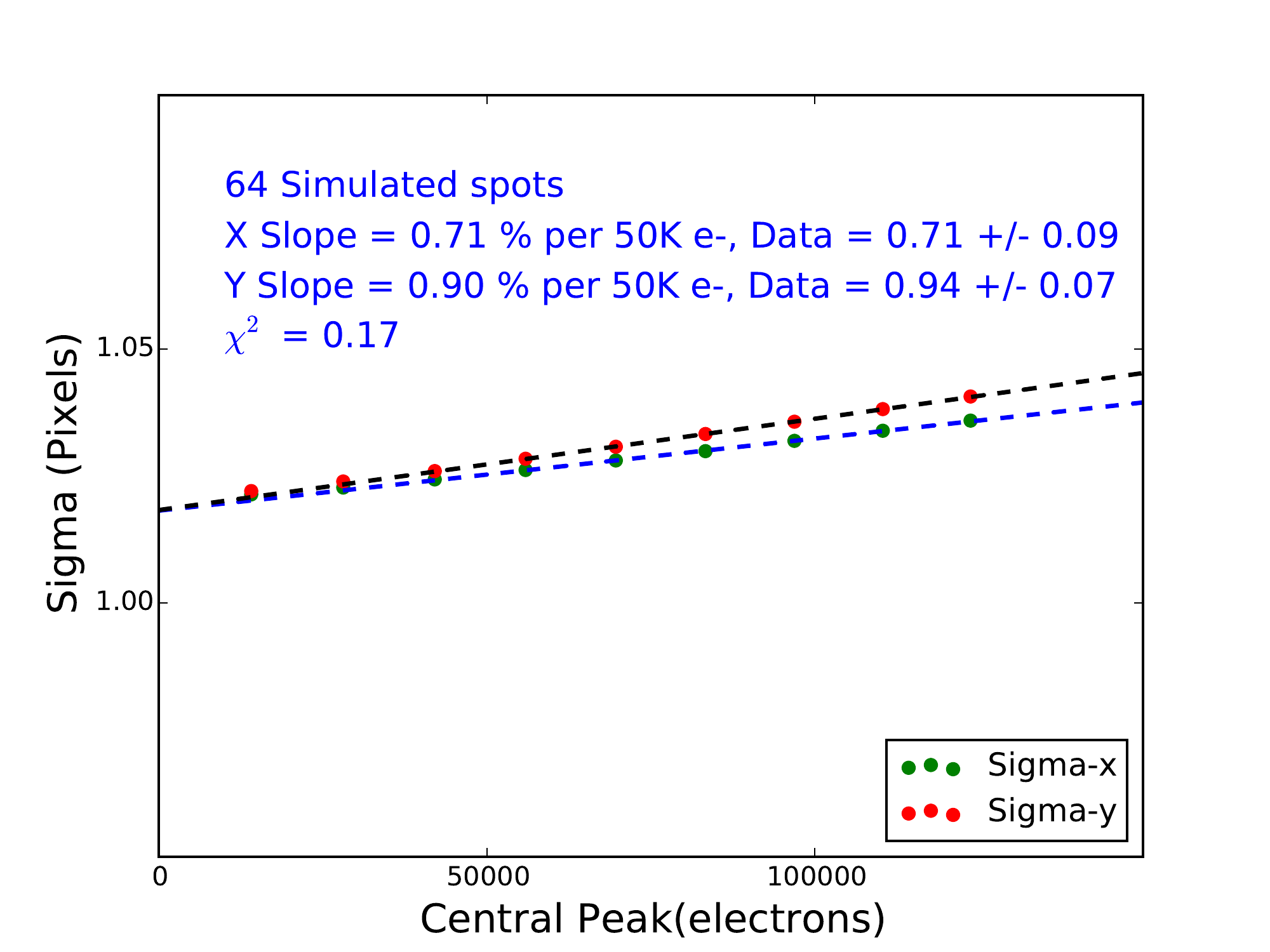}}
  \subfigure{\includegraphics[trim=0.2in 0.0in 0.5in 0.6in,clip,width=0.495\textwidth]{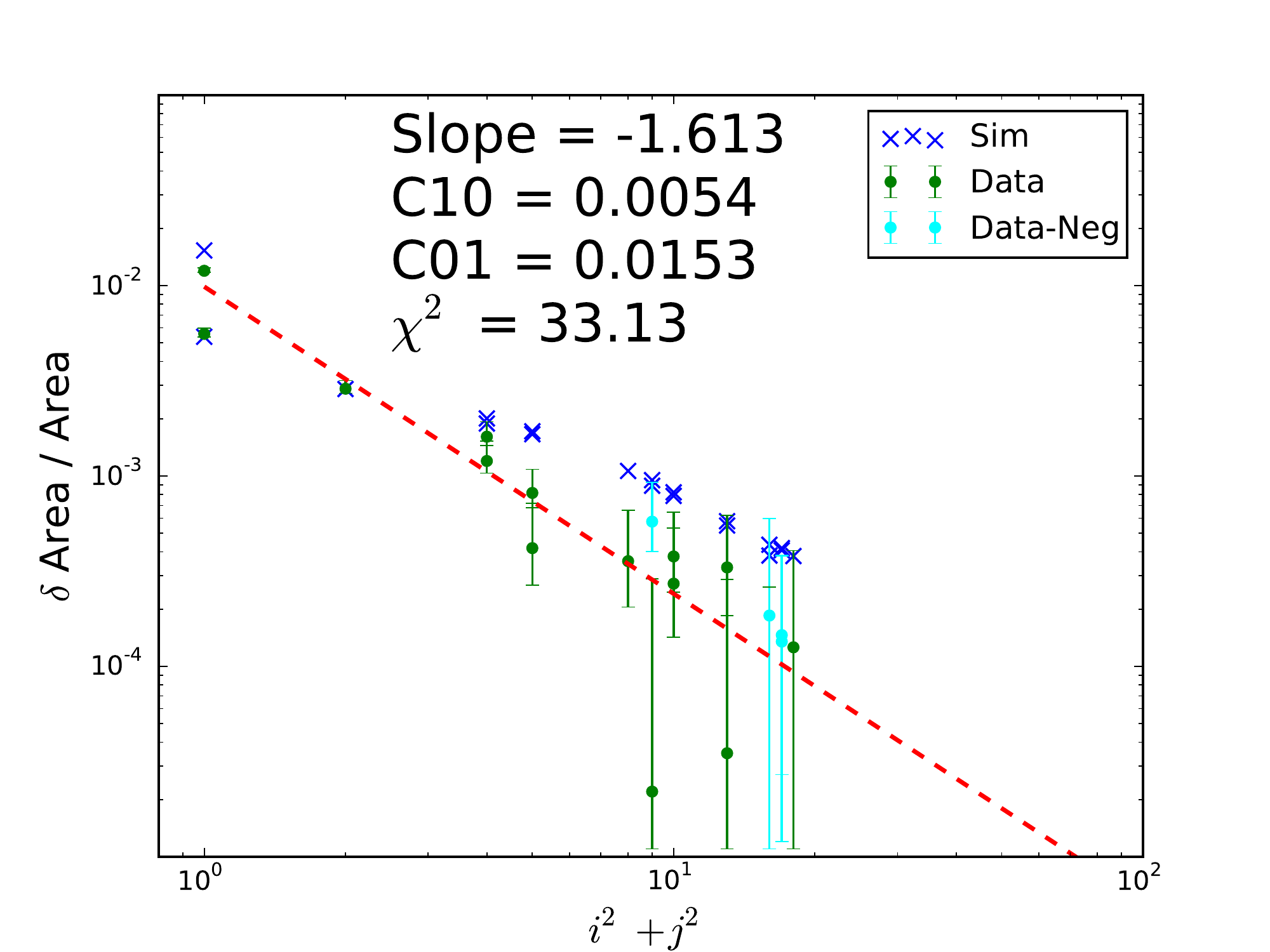}}
  \subfigure{\includegraphics[trim=0.2in 0.1in 0.5in 0.6in,clip,width=0.495\textwidth]{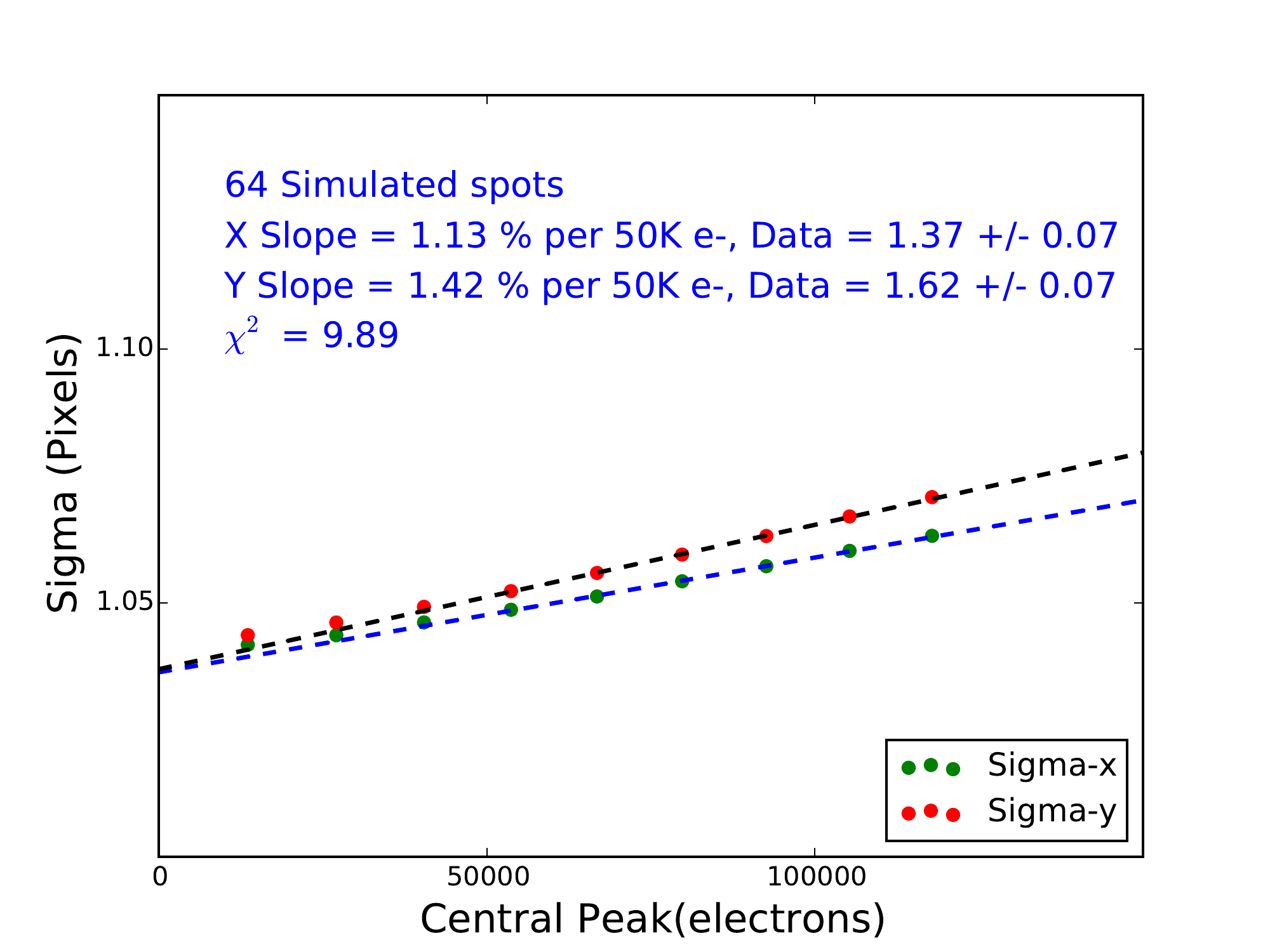}}
  \subfigure{\includegraphics[trim=0.2in 0.0in 0.5in 0.6in,clip,width=0.495\textwidth]{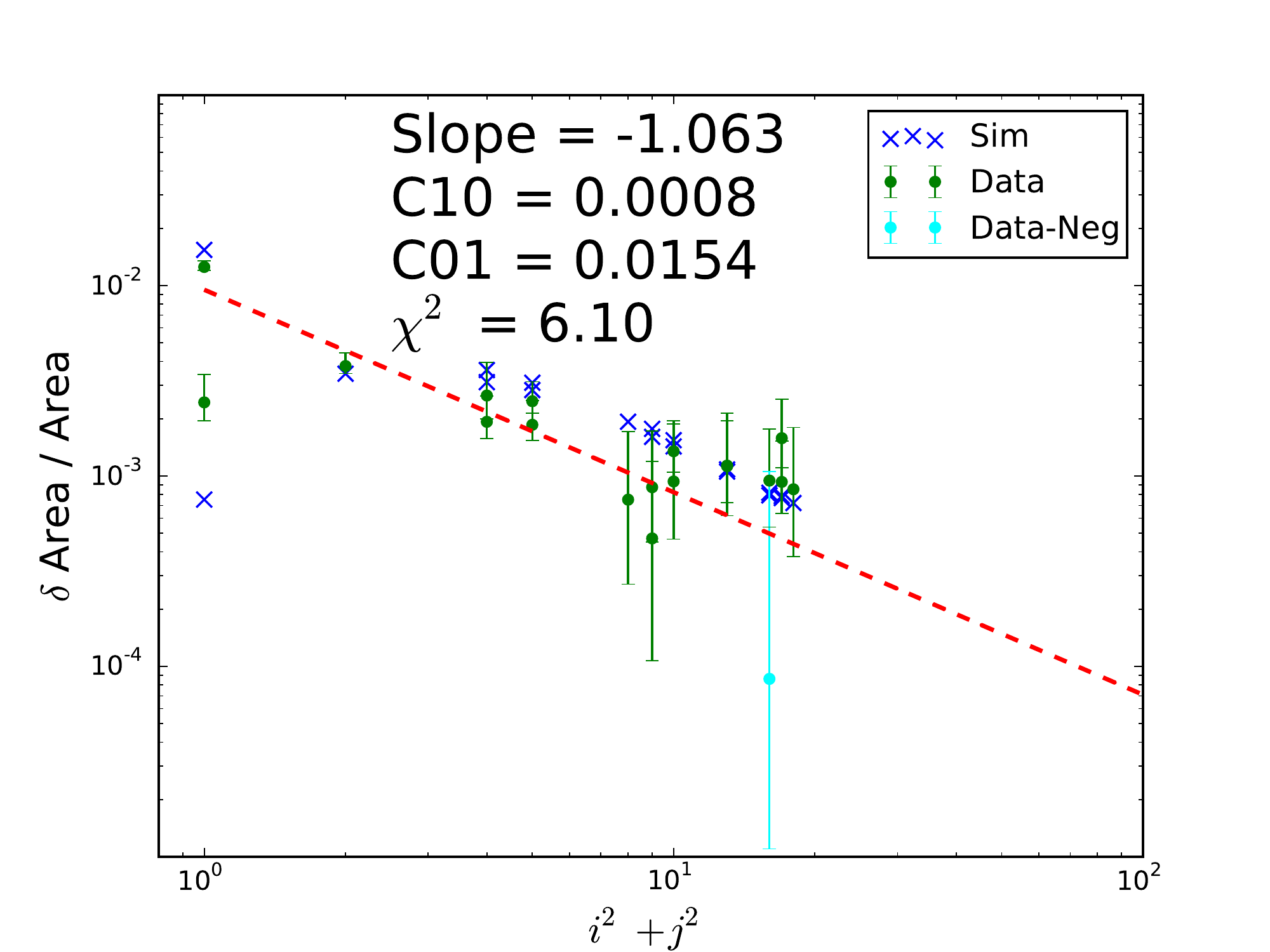}}
  \subfigure{\includegraphics[trim=0.2in 0.1in 0.5in 0.6in,clip,width=0.495\textwidth]{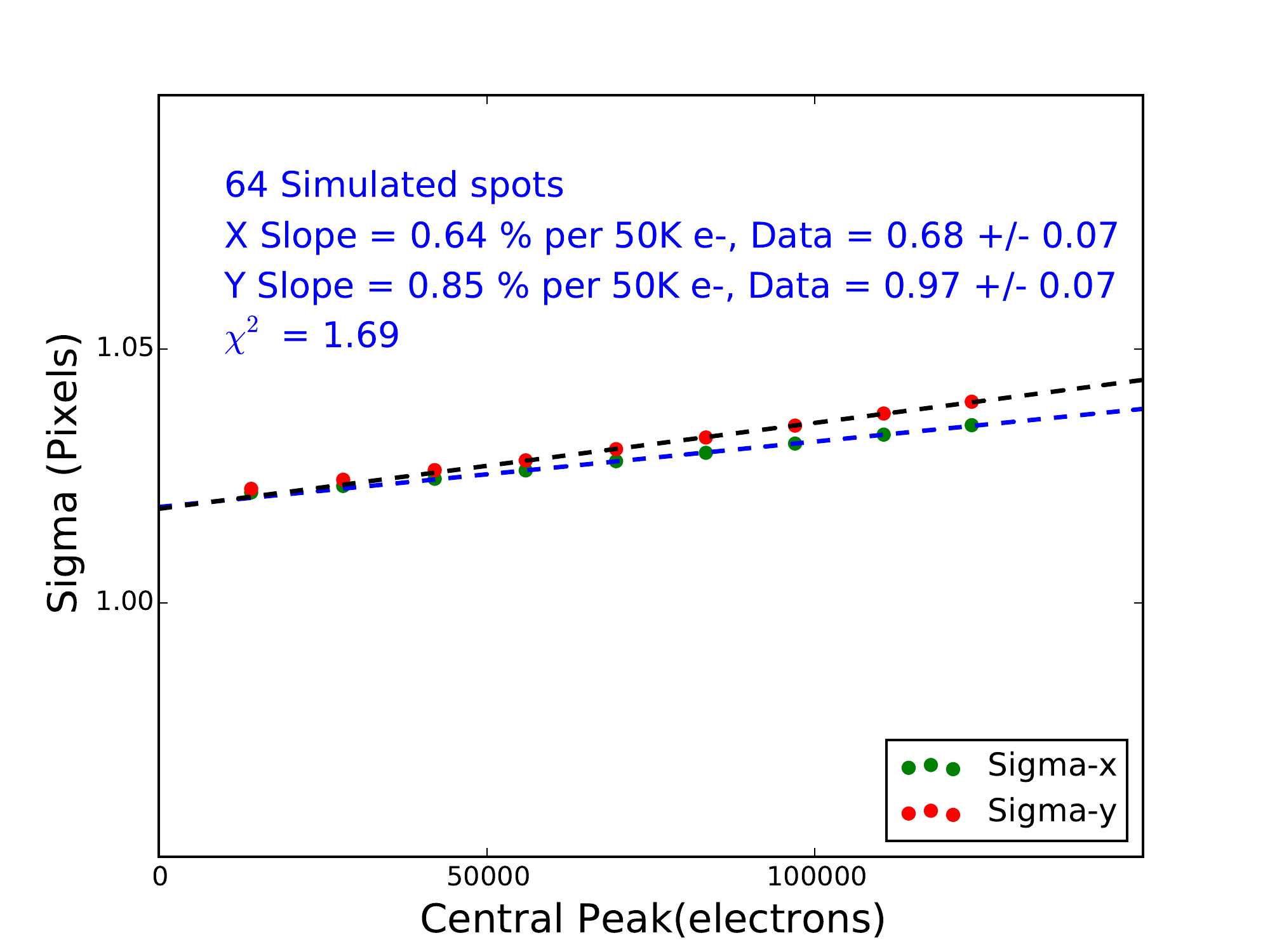}}
  \subfigure{\includegraphics[trim=0.2in 0.0in 0.5in 0.6in,clip,width=0.495\textwidth]{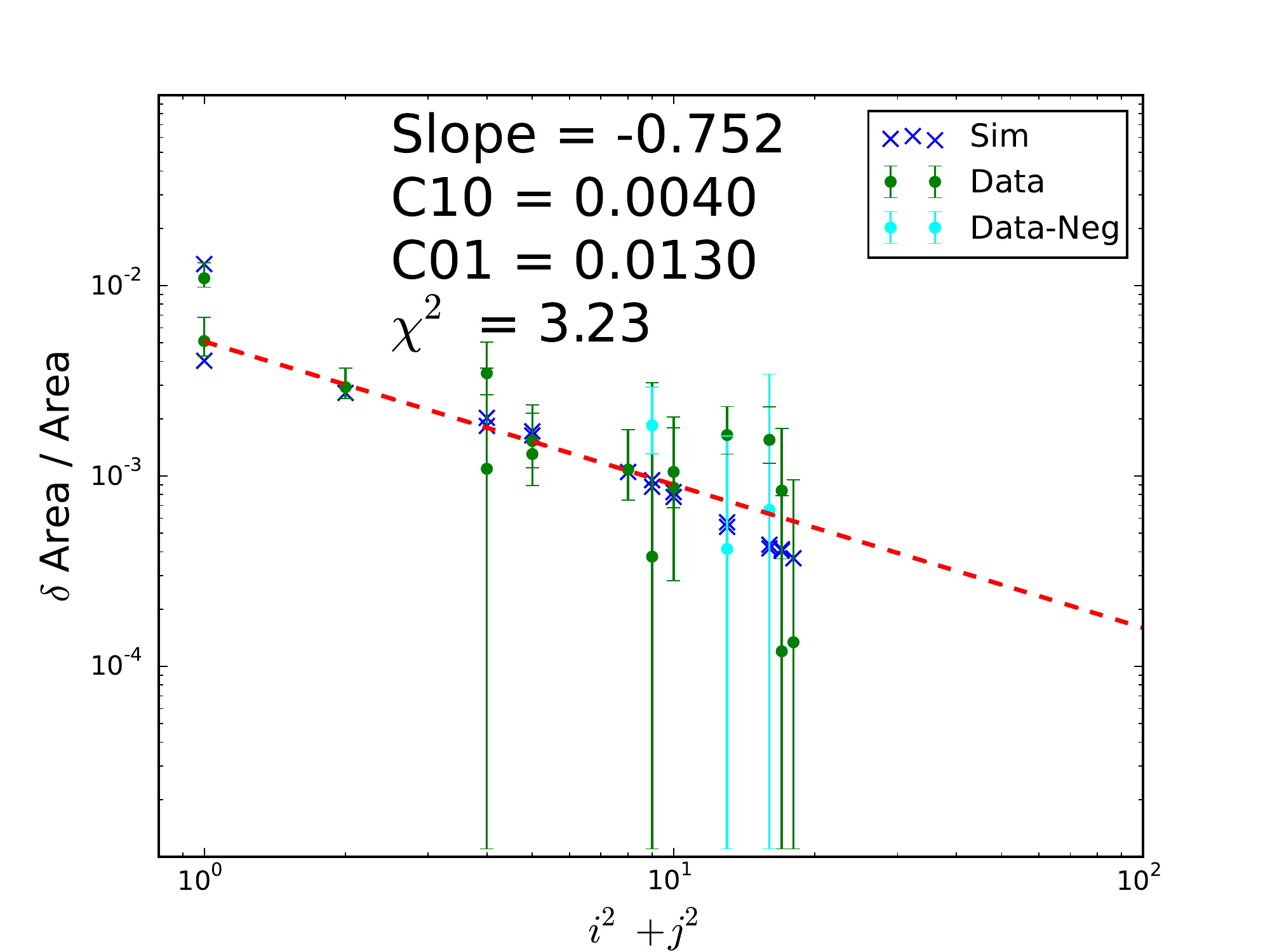}}
  \caption{Measurements and Simulations.  Top row: Vbb = -60V, Vp = +4/-8V.  Middle row: Vbb = -30V, Vp = +4/-8V.  Bottom row: Vbb = -60V, Vp = +6/-6V.  Left column: The ``Data'' lines for the baseline condition (top plot) consist of 9 separate measurements, while the ``Data'' lines for the bottom two plots consist of 5 separate measurements each.  The simulations are correctly capturing the B-F slopes as a function of voltage conditions, as well as capturing the X-Y asymmetry.  Right column:  The measurements for the baseline condition (top plot) consists of 2000 flats, while the data for the bottom two plots consist of 500 flats each.  The pixel-pixel correlations are also well fit by the simulations.  The simulations of direct spot measurements and pixel-pixel correlations are all run with the same code and with the same CCD model parameters.  Consistently fitting both sets of measurements with a single model is confirmation of the validity of the model.}
  \label{Meas_vs_Sim}
\end{figure}

\appendix
\section{Appendix - Relation of Pixel-Pixel Correlations to Pixel Area Change}
\renewcommand{\thefigure}{A.\arabic{figure}}
\setcounter{figure}{0}
\label{Correlations_Appendix}

The purpose of this appendix is to clarify the relationship between the pixel-pixel correlations and the pixel area changes caused by the lateral electric fields which distort the pixel boundaries.  It is clear that the two are related, but it is not immediately obvious whether the covariance of a given pixel is equal to the fractional change in area, proportional to it, or whether they are related in some non-linear way.  To clarify this, this appendix discusses a simple toy model which has been both analyzed analytically and simulated numerically.
\begin{figure}[H]
  \begin{center}
  \includegraphics[trim=3.0in 4.0in 3.0in 1.1in,clip,width=0.35\textwidth]{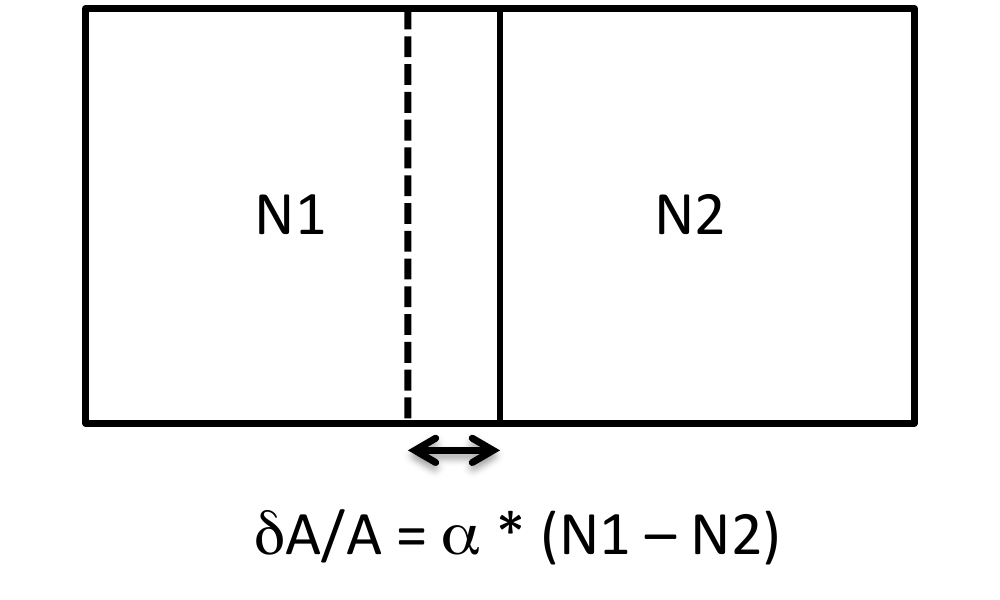}
  \caption{Toy Model of the B-F effect with two pixels with a movable partition between them.}
  \end{center}
  \label{Toy_Model}
\end{figure}

\begin{figure}[H]
  \centering
  \subfigure[10K e-/pixel; 100K trials]{\includegraphics[trim=0.0in 0.0in 0.0in 0.0in,clip,width=0.45\textwidth]{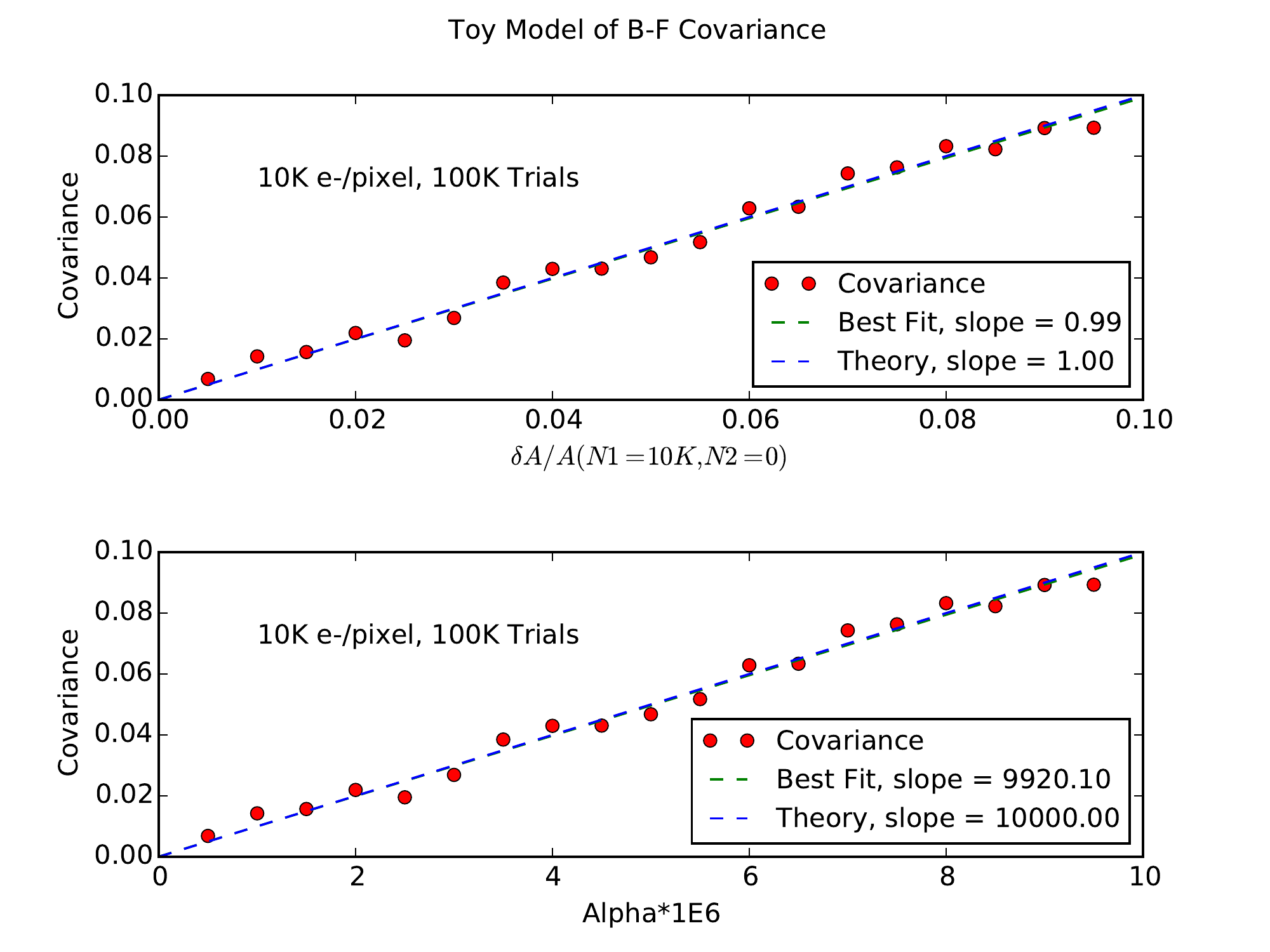}}
  \subfigure[50K e-/pixel; 100K trials]{\includegraphics[trim=0.0in 0.0in 0.0in 0.0in,clip,width=0.45\textwidth]{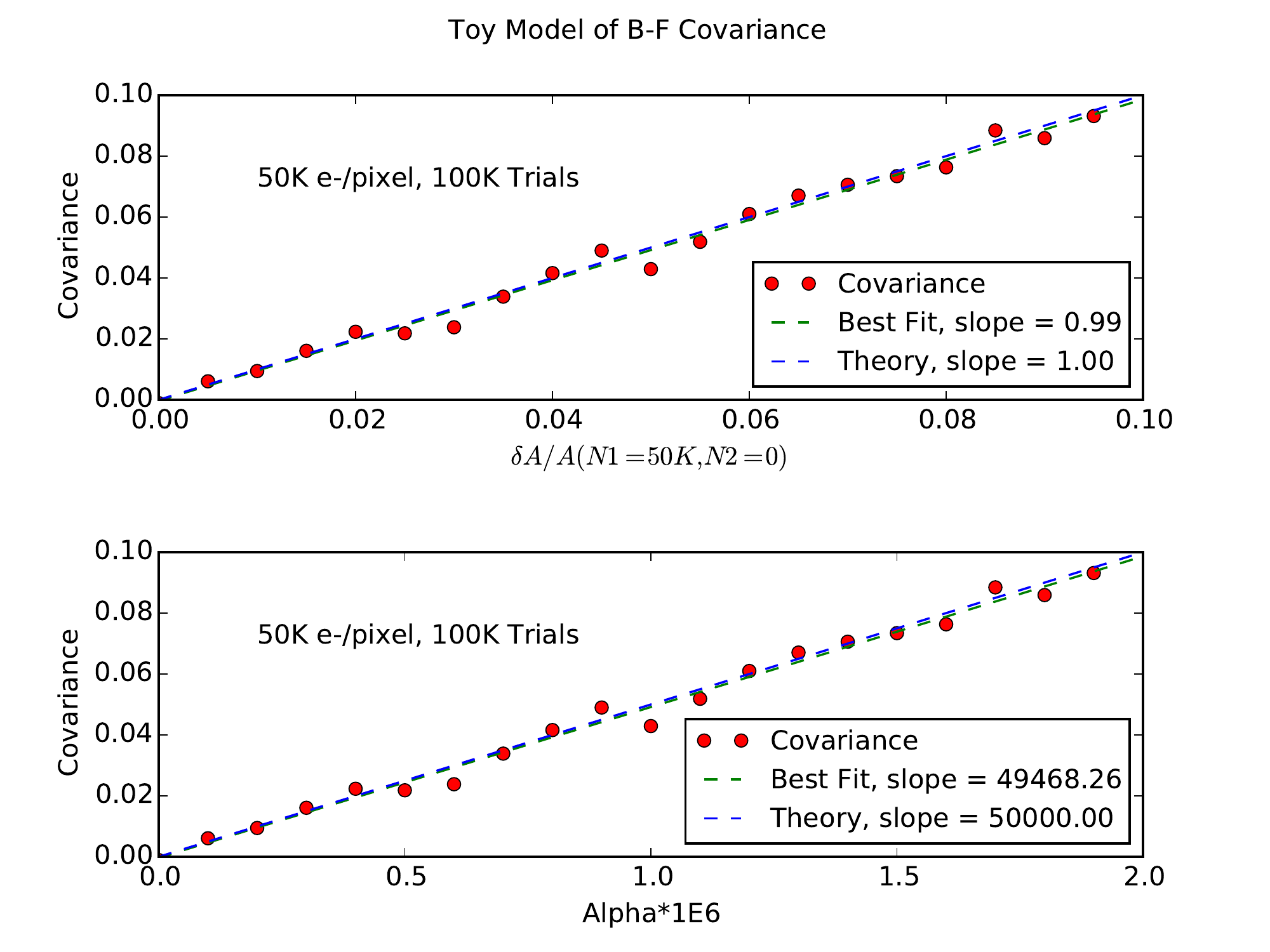}}
  \caption{Monte-Carlo simulations of the toy model.}
  \label{Toy_Sim}
\end{figure}

The basics of the toy model are shown in Figure \ref{Toy_Model}.1.  There are two initially equal pixels with a movable partition between them.  Each pixel has a number of stored electrons ($\rm N_1, N_2$), and the partition is assumed to move by an amount proportional to the difference in stored charge between the two pixels, as shown in the figure.  We assume that the collection of charge in the two pixels is a Poisson process with rate parameter $\lambda$, so that the probability of collecting N electrons is:
\begin{equation}
\rm P(N; \lambda) = \frac{\lambda^N e^\lambda}{N!}
\end{equation}
As is well known, a Poisson process can be decomposed into two subprocesses as follows:
  \begin{equation}
\rm P(N_1, N_2; \lambda_1, \lambda_2) = \frac{\lambda_1^{N_1} e^{\lambda_1}}{N_1!} \frac{\lambda_2^{N_2} e^{\lambda_2}}{N_2!}
  \end{equation}
  In this case, because the rate of collecting electrons is proportional to the pixel area, the two rate parameters ($\rm \lambda_1, \lambda_2$) are related to the global parameter $\rm \lambda_0$ as follows:
  \begin{equation}
\rm \lambda_1 = \lambda_0(1 - \frac{\alpha}{2}(N_1-N_2)); \lambda_2 = \lambda_0(1 + \frac{\alpha}{2}(N_1-N_2));
  \end{equation}
The factor of 2 in the denominator is inserted because the initial areas are equal, and the final areas are determined by the final counts ($\rm N_1, N_2$). Integrating from an initial value of $\delta A = 0$ to the final value $\delta A = A \alpha(N_1 - N_2)$ gives this factor of two.  Now we insert $\lambda_1, \lambda_2$ into P and expand, keeping terms to first order in $\alpha$:
  \begin{equation}
\rm P(N_1, N_2; \lambda_0, \alpha) = P(N_1; \lambda_0) P(N_2; \lambda_0) (1 - \frac{\alpha}{2}(N_1-N_2)^2)
  \end{equation}
The means, variances, and covariance are then (to first order in $\alpha$):
  \begin{equation}
\rm Mean(N_1) = \sum_{N_1 = 0}^\infty \sum_{N_2 = 0}^\infty N_1 P(N_1, N_2; \lambda_0, \alpha) = \lambda_0 (1 - \frac{\alpha}{2})
  \end{equation}
  \begin{equation}
\rm Var(N_1) = \sum_{N_1 = 0}^\infty \sum_{N_2 = 0}^\infty (N_1 - Mean(N_1) )^2 P(N_1, N_2; \lambda_0, \alpha) = \lambda_0 (1 - \frac{\alpha}{2}(\lambda_0 + \frac{1}{2}))
  \end{equation}
  \begin{equation}
\rm Correlation(N_1, N_2) = \sum_{N_1 = 0}^\infty \sum_{N_2 = 0}^\infty (N_1 - Mean(N_1) )(N_2-Mean(N_2) ) P(N_1, N_2; \lambda_0, \alpha) = \lambda_0^2 \alpha
  \end{equation}
  \begin{equation}
    \rm Covariance(N_1, N_2) = \frac{Correlation(N_1, N_2)}{Var(N_1)} = \lambda_0 \alpha
    \label{Cov_Eq}
  \end{equation}
When calculating the changes in area as in Figure \ref{Pixel_Distortion}, we set $\rm N_1 = N_0, N_2 = 0$, so that $\rm \frac{\delta A}{A} = \alpha N_0 = \alpha \lambda_0$.  So we find that the fractional change in area of a pixel and the covariance of that pixel are in fact equal. 
As a further check, we have built a Monte-Carlo version of the toy model, simply ``throwing darts'' at the two pixels using a pseudo-random number generator, and moving the central partition after each ``dart''.  The results of running the Monte-Carlo code for 20 different values of $\rm \alpha$ and two different values of $\rm \lambda_0$ are show in Figure \ref{Toy_Sim}, and confirm Equation \ref{Cov_Eq}.

The results of the toy model can be extended to two dimensions, and the fact that the right column of Figure \ref{Meas_vs_Sim} agrees with the measured data confirms the reasoning.

\bibliographystyle{unsrt}
\bibliography{ccd}

\end{document}